\documentclass[aps,prd,preprint,preprintnumbers,unsortedaddress,superscriptaddress,showpacs,nofootinbib]{revtex4-1}

\pdfoutput=1
\usepackage{bm}
\usepackage{graphicx}
\usepackage{amsmath}
\usepackage{amsfonts}
\usepackage{amssymb}
\usepackage{color}
\usepackage{ulem}

\usepackage[section]{placeins}

\usepackage{booktabs}
\usepackage[colorlinks, citecolor=blue,anchorcolor=red,menucolor=red, linkcolor=red,filecolor=red,runcolor=red,urlcolor=blue,frenchlinks=red]{hyperref}

\newcommand{\roca}[1]{\textcolor{black}{#1}}

\newcommand{\nn}{\nonumber}

\begin{document}
\bibliographystyle{unsrt}
\arraycolsep1.5pt

\title{$\tau$ decay into a pseudoscalar and an axial-vector meson}

\author{L.~R.~Dai}
\email{dailr@lnnu.edu.cn}
\affiliation{Department of Physics, Liaoning Normal University, Dalian 116029, China}
\affiliation{Departamento de F\'isica Te\'orica and IFIC, Centro Mixto Universidad de Valencia-CSIC,
Institutos de Investigac\'ion de Paterna, Aptdo. 22085, 46071 Valencia, Spain
}

\author{L. Roca}
\email{luisroca@um.es}
\affiliation{Departamento de F\'isica, Universidad de Murcia, E-30100 Murcia, Spain}

\author{E.~Oset}
\email{oset@ific.uv.es}
\affiliation{Departamento de F\'isica Te\'orica and IFIC, Centro Mixto Universidad de Valencia-CSIC,
Institutos de Investigac\'ion de Paterna, Aptdo. 22085, 46071 Valencia, Spain
}

\date{\today}
\begin{abstract}

We study theoretically the decay $\tau^- \to  \nu_\tau P^- A$, with $P^-$ a $\pi^-$ or $K^-$ and $A$ an axial-vector resonance $b_1(1235)$, $h_1(1170)$,
$h_1(1380)$, $a_1(1260)$, $f_1(1285)$ or any of the two  poles of
the $K_1(1270)$. The process proceeds through a triangle mechanism where a vector meson pair is first produced from the weak current and then one of the vectors produces two pseudoscalars, one of which reinteracts with the other vector to produce the axial resonance. For the initial weak hadronic production we use a recent formalism to account for the hadronization after the initial quark-antiquark pair produced from the weak current, which explicitly filters G-parity states and obtain easy analytic formulas after working out the angular momentum algebra. The model also takes advantage of the chiral unitary theories to evaluate the vector-pseudoscalar amplitudes, where the axial-vector resonances were obtained as dynamically generated from the VP interaction. We make predictions for invariant mass distribution and branching ratios for the channels considered.

\end{abstract}

\maketitle

\section{Introduction} \label{sec:intro}

\roca{
The fact that the $\tau$ is the only lepton heavy enough to decay into hadrons makes  the hadronic $\tau$ lepton decays a priceless test of the strong interaction at low energy in the light flavor sector \cite{Schael:2005am,Davier:2005xq,Braaten:1991qm,Lafferty:2015hja}. The intermediate and final state decay hadrons are usually produced with lower background than in other low energy processes.
Even though there are more than 100 hadronic  $\tau$ decays experimentally reported by the PDG \cite{pdg} (which account for about 65\% of the $\tau$  decay width ), it is also clear that not all the possible ones have been observed or whether there is no room for decays beyond the standard model.
While inclusive reactions are well suited for accurate extraction of standard model parameters such as the strong coupling constant \cite{Braaten:1991qm,Boito:2014sta,Pich:2013lsa}, the exclusive ones are much more involved and difficult to predict within  QCD, and here is where effective theories for hadronic low energy interactions gain prominence.
 Special theoretical attention
 has been devoted to decay channels with two and three pseudoscalar mesons in the final state (see \cite{Portoles:2007cx} for a brief review). Channels with more pseudoscalars or other mesons like vector or axial-vector ones are less studied \cite{Schael:2005am,Davier:2013sfa}.
Particularly, very poorly understood are the channels with one pseudoscalar plus one axial-vector meson in the final state, which are the aim of the study in the present work. Experimentally only the  $f_1(1285)\pi$ channel has been measured \cite{pdg}.
It is at this point where effective theories of strong interactions at low energies can stand up. Particularly, the unitary extensions of chiral perturbation theory (U$\chi$PT) provide a dynamical and powerful explanation of the generation of the low-lying axial-vector resonances \cite{Lutz:2003fm,roca05,Zhou:2014ila}. With the only input of the lowest order chiral perturbation theory Lagrangians and the implementation of unitarity in coupled channels, most of the lowest mass axial vector resonances
($b_1(1235)$, $h_1(1170)$,
$h_1(1380)$, $a_1(1260)$, $f_1(1285)$ and  two poles
for the $K_1(1270)$)
were dynamically obtained \cite{Lutz:2003fm,roca05,Zhou:2014ila} as poles in the pseudoscalar-vector (PV) scattering amplitudes, without the need to include them as explicit degrees of freedom. With only one free parameter (for regularization of PV loops), this model predicts not only masses and widths of these axial vector resonances but also the full PV scattering amplitudes from where {\it{e.g.}} the coupling of the different resonances to the different PV channels can be obtained.
Within this model, the $\tau$ decay into one pseudoscalar plus one axial-vector resonance  requires the production of one pseudoscalar and one PV pair in the hadronization process, since these axial-vector resonances are dynamically generated from the PV interaction.  This can be dominantly accomplished via a triangular mechanism of the kind shown in Fig.~\ref{fig:diane1}. Actually, for the $f_1(1285)\pi$ case, it was shown in \cite{Oset:2018zgc} that it was the dominant contribution. Indeed, for some particular kinematic conditions, the triangle diagram  benefits from a large enhancement since it develops a singularity \cite{landau}, which according to the Coleman Norton theorem \cite{ncol} is related to the classical process in which a particle decays into two particles A and B, then A decays into two other particles, and one of them merges with particle B to produce a third particle. A new
reformulation  of these findings  can be seen in  \cite{Bayar:2016ftu,Schmid}.  A similar mechanism has been also recently used for the decay $\tau^-\to \nu_\tau \pi^- f_0(980) (a_0(980))$ \cite{Dai:2018rra}.}

\roca{
On the other hand, for the hadronization process from the $W^-$ boson to two mesons, we follow the approach of  \cite{tdai}, where the $^3P_0$ model \cite{LeYaouanc:1972vsx,close3P0} is used
to hadronize the primary quarks produced from the weak interaction,  working out all the angular momentum and spin algebra.
In particular, a Cabibbo favored $d\bar u$ pair is produced from the $W^-$ which then hadronizes from an extra $q\bar q$ pair with the quantum numbers of the vacuum. The different possible final meson-meson states are related by $SU(3)$ symmetry.
The strength of the formalism in \cite{tdai} is that it
carries on an elaborate calculation of the angular momentum and spin algebra which allows, in the end, to rely upon only one global unknown constant to get all the different channels.
This unknown factor is obtained in the present work from the experimental value of the
$\tau^-\to \nu_\tau \pi^- K^* \bar K^*$ decay.
Furthermore, the formalism allows
for an explicit filter of G-parity states, of special importance in the present work.}

\section{Formalism}

\subsection{Feynman diagrams}

\begin{figure}[ht!]
\includegraphics[scale=0.65]{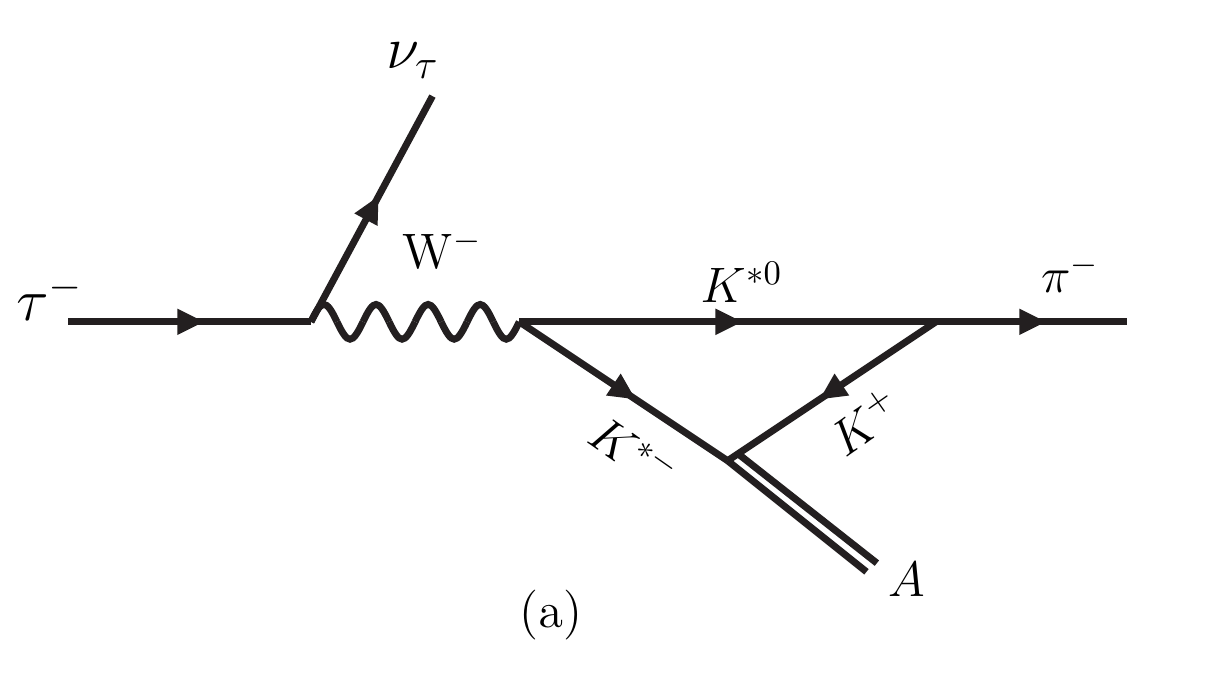} \includegraphics[scale=0.65]{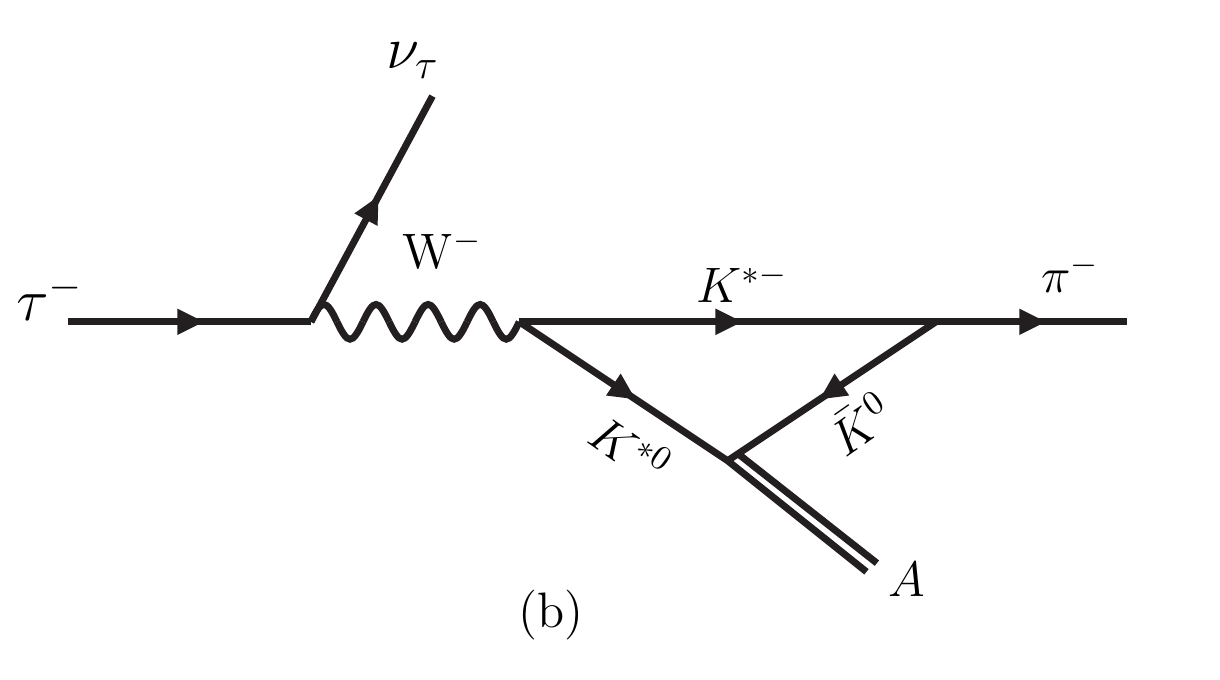}
\caption{Diagrams for the decay of $\tau^- \to \nu_\tau \pi^- A$, with $A$  axial vectors.}
\label{fig:diane1}
\end{figure}

\begin{figure}
	\begin{center}
		\includegraphics[width=0.45\textwidth]{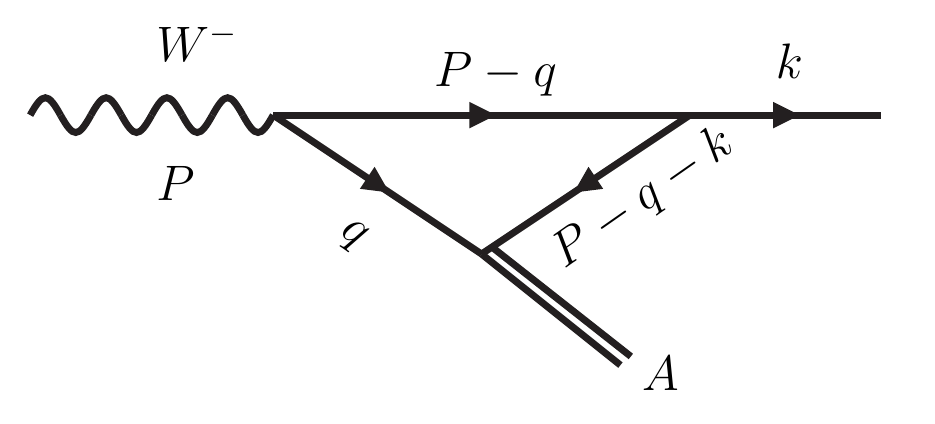}
	\end{center}
	\caption{\label{fig:mom} The momenta assignment for the decay process}
\end{figure}

\roca{We are going to study the decays} $\tau^- \to \nu_\tau \pi^- A $, with  $A$ being axial-vector resonances,
including the positive $G$-parity $f_1(1285)$, $b_1(1235)$ states, and  negative $G$-parity $h_1(1170)$, $h_1(1380)$ and  $a_1(1260)$ states,
and the $\tau^- \to \nu_\tau K^- K_1(1270)$ decay. \roca{ As mentioned in the Introduction, these were the low-lying axial-vector resonances  dynamically generated in \cite{roca05}. Actually, for the $K_1(1270)$ resonance, it was shown in \cite{roca05,geng07} that is has a two pole structure and then we will consider both of them. Since the A resonances are dynamically generated from the PV interaction, for the  $\tau^- \to \nu_\tau P^- A$, the way to produce  the VP to generate axials and the extra pseudoscalar in the final state is via a triangular mechanism of the kind shown in fig.~\ref{fig:diane1}. The well defined G-parity axial-vector states have dominant couplings to either $\rho\pi$ or $K^* \bar K$ \cite{roca05}.
Therefore,} for the  positive  $G$-parity \roca{axial-vector} states \roca{ ($f_1(1285)$ and  $b_1(1235)$)}, the complete Feynman diagrams for the decay with the triangle mechanism  are \roca{those} shown in Fig.~\ref{fig:diane1}.
Fig. \ref{fig:diane1}(a)
shows the process $\tau^- \to \nu_\tau K^{*0} K^{*-}$ followed by the $K^{*0}$ decay into $\pi^- K^+$  and the merging of the $K^{*-} K^+$ into $A$;
and Fig.\ref{fig:diane1}(b) shows  the process $\tau^- \to \nu_\tau K^{*-} K^0$  followed by the $K^{*-} $ decay into $\pi^-\bar{K}^0$ and the merging of the $ K^{*0} \bar{K}^0$ into $A$.
The momenta assignment for the decay process is given in Fig.~\ref{fig:mom}.
The \roca{needed} couplings obtained in \cite{roca05,Oset:2018zgc} for positive $G$-parity  \roca{axial-vector} states to the appropriate G-parity $VP$ eigenstates  are given in  Table~\ref{tab:popole}.

\begin{table}[h!]
\footnotesize
\begin{center}
\caption{ Couplings for positive  $G$-parity  states \cite{roca05,Oset:2018zgc} (All the units are MeV).}
\begin{tabular}{c|cc|cc}
\hline
 & \multicolumn{2}{c|}{~~~~~~$f_1(1285)$~~~~~~} &\multicolumn{2}{c}{~~~~~~ $b_1(1235)$~ ~~~~~~} \\\cline{2-5}
couplings& $g_i$ & $|g_i|$ & $g_i$ & ~~$|g_i|$~~ \\
\hline &&&& \\[-6.5mm]
$\frac{1}{\sqrt{2}}(\bar K^*K - K^*\bar K)$    &~~  \roca{$7350+i0$} ~~&~~  \roca{$7350$} ~~ &~~ $-$ & $-$ ~  \\
$\frac{1}{\sqrt{2}}(\bar K^*K + K^*\bar K)$  & $-$& $-$ & ~~ $6172-i75$ ~~&~~ $6172$  \\
\hline
\end{tabular}
\label{tab:popole}
\end{center}
\end{table}

\begin{figure}[ht!]
\includegraphics[scale=0.65]{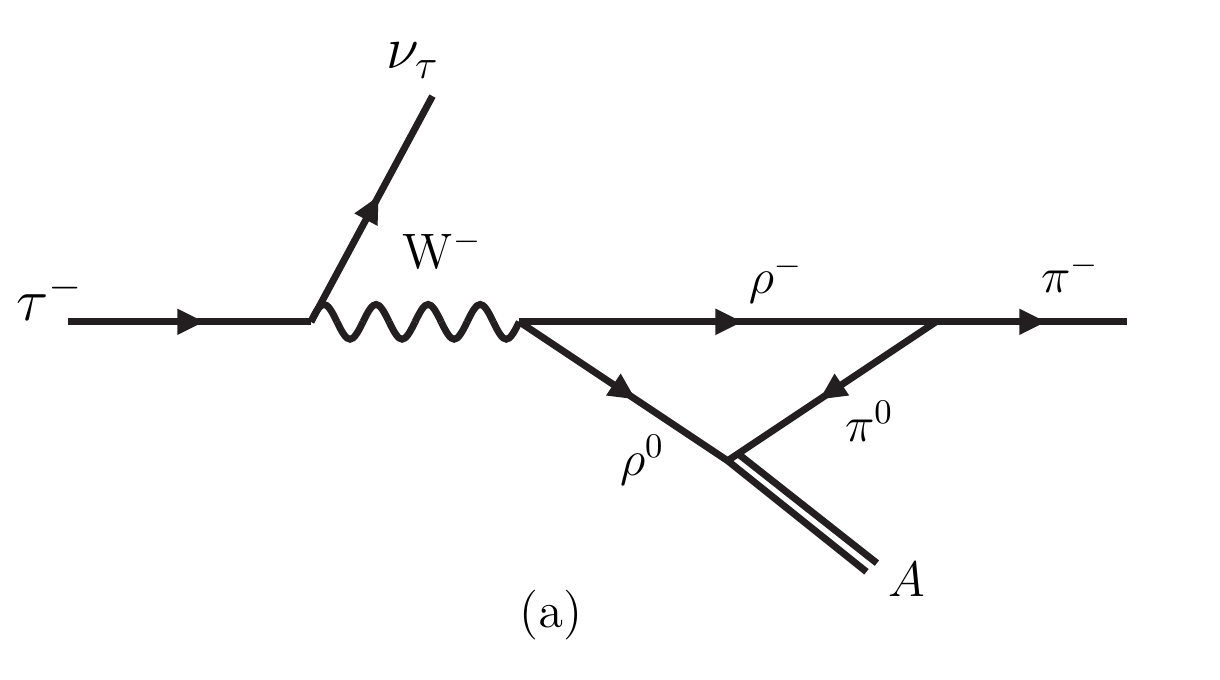} \includegraphics[scale=0.65]{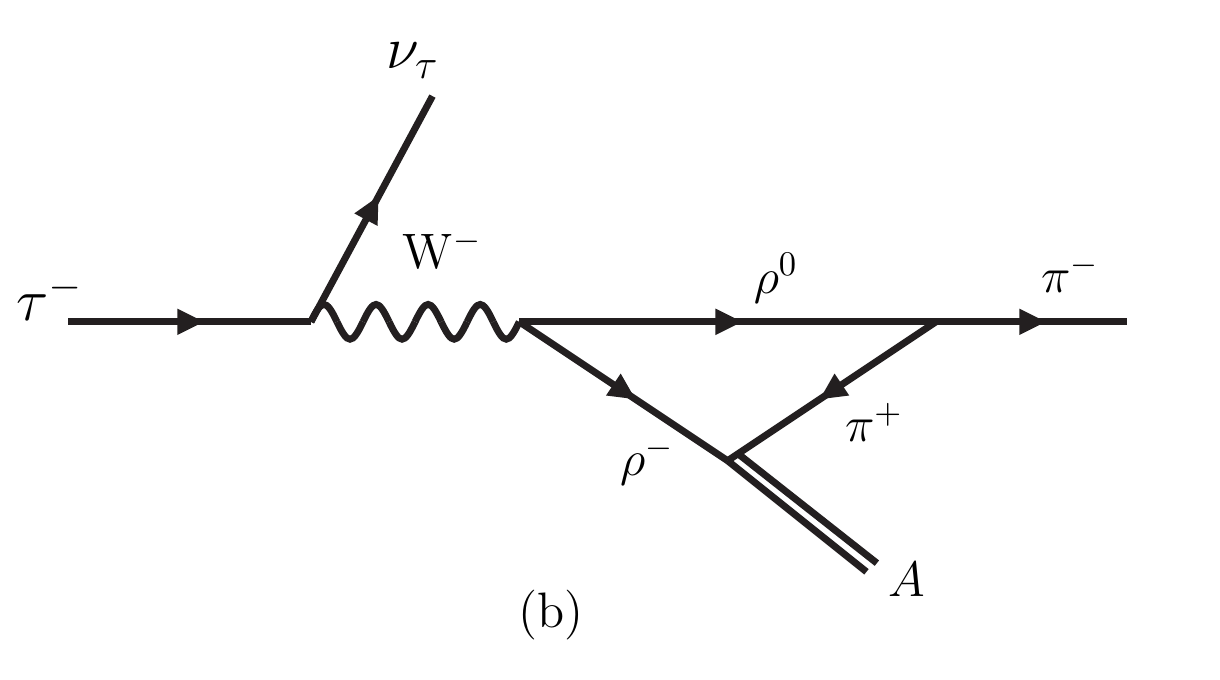}
\caption{The same as Fig. \ref{fig:diane1} but for $\tau^- \to \nu_\tau \rho^{-} \rho^0,  \nu_\tau \rho^{0} \rho^-$ decays.}
\label{fig:diane2}
\end{figure}

For the negative $G$-parity  \roca{axial-vector} states \roca{( $h_1(1170)$,  $h_1(1380)$ and $a_1(1260)$)},  we need \roca{ to consider the diagrams in  Fig. \ref{fig:diane2} in addition to those in
Fig. \ref{fig:diane1} because $\rho \rho$ has $G=+$ and,  the fact that $G(\pi)=-$ demands that the axial has negative $G-$parity to have G-parity conservation. This mechanism with initial $\rho\rho$ production is thus not possible for the  positive $G$-parity axials $f_1(1285)$ and  $b_1(1235)$.}
 Fig. \ref{fig:diane2} (a) shows  the process  $\tau^- \to \nu_\tau \rho^{-} \rho^0$  followed by the $\rho^{-} $ decay into $\pi^-\pi^0$ and the merging of the $ \rho^0 \pi^0$ into $A$;
and Fig. \ref{fig:diane2}(b) shows  the process $\tau^- \to \nu_\tau \rho^{0} \rho^-$   followed by the $\rho^{0} $ decay into $\pi^-\pi^+$ and the merging of the $ \rho^- \pi^+$
into $A$.  The  \roca{needed} couplings \cite{roca05} for negative $G$-parity  states are given in Table \ref{tab:nepole}.
\begin{table}[h!]
\begin{center}
\caption{The same as Table \ref{tab:popole} but for negative $G$-parity  states.}
\begin{tabular}{c|cc|cc|cc}
\hline
   & \multicolumn{2}{c|}{$h_1(1170)$} & \multicolumn{2}{c|}{ $h_1(1380)$} & \multicolumn{2}{c}{ $a_1(1260)$}\\
\cline{2-7} &&&&&& \\[-6.5mm]
couplings & $g_i$ & $|g_i|$ & $g_i$ & $|g_i|$& $g_i$ & $|g_i|$ \\
\hline &&&&&& \\[-6.5mm]
$\rho\pi$ & $-3453+i1681$ & $3840$~~& ~~$648-i959$& $1157$ & $-3795+i2330$ & $4453$ \\
$\frac{1}{\sqrt{2}}(\bar K^*K+K^*\bar K)$  & $781-i498$ & $926$ & $6147+i183$ & $6150$ &$-$&$-$\\
$\frac{1}{\sqrt{2}}(\bar K^*K -K^*\bar K)$  & $-$ & $-$ & $-$ & $-$ &$1872-i1486$&$2390$\\
\hline
\end{tabular}\label{tab:nepole}
\end{center}
\end{table}

\begin{figure}[ht]
\includegraphics[scale=0.68]{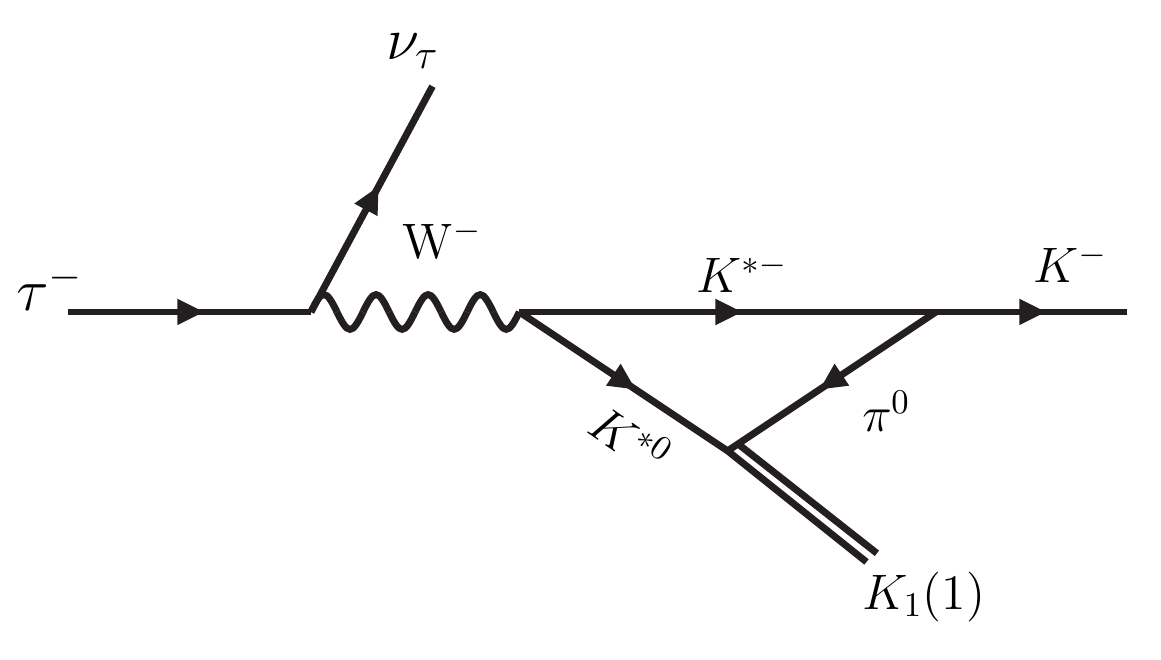}
\caption{ Diagram for the decay of  $\tau^- \to \nu_\tau K^- K_{1} (1) )$, where $K_{1} (1)$ is the first pole of  $K_{1} (1270)$.}
\label{fig:diak1}
\end{figure}
\begin{figure}[ht]
\includegraphics[scale=0.65]{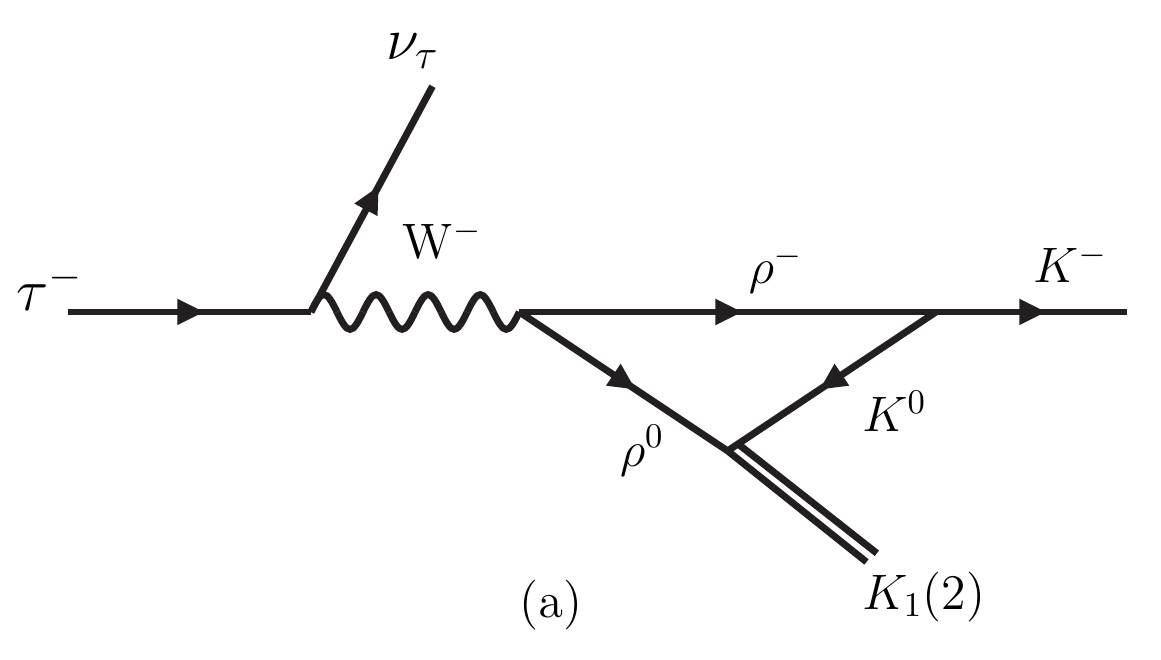} \includegraphics[scale=0.65]{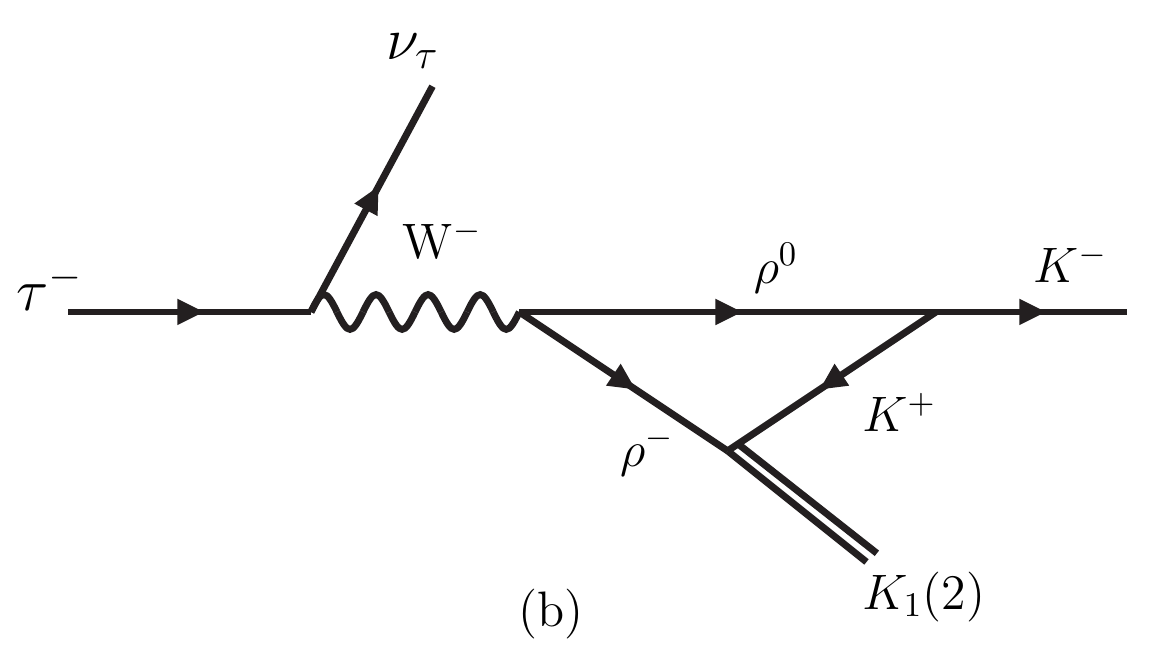}
\caption{ Diagram for the decay of  $\tau^- \to \nu_\tau K^- K_{1} (2) )$, where $K_{1} (2)$ is the second  pole of  $K_{1} (1270)$.}
\label{fig:diakx1}
\end{figure}

\roca{In  \cite{roca05,geng07} two poles for the $K_1(1270)$ where obtained at complex energy  positions $1195-i123$~MeV and $1284-i73$~MeV in unphysical Riemann sheets. The lowest mass pole, which we will call in the following $K_1(1)$, couples mostly to $K^*\pi$ and the highest one, $K_1(2)$, to $\rho K$. The dominant couplings are shown in  Table \ref{tab:k1pole} \cite{geng07}.}

\begin{table}[h!]
\begin{center}
\caption{The same as Table \ref{tab:nepole} but for two $K_1(1270)$ states.}
\begin{tabular}{c|cc|cc}
\hline &&&& \\[-6.5mm]
 & \multicolumn{2}{c|}{$K_{1} (1)$} &\multicolumn{2}{c}{ $K_{1} (2)$ } \\
 \cline{2-5}
couplings & $g_i$ & $|g_i|$ & $g_i$ & $|g_i|$ \\
\hline  &&&& \\[-6.5mm]
$\rho K$   &  $-$& $-$ &~~~  $5274+i297$ ~~~  & ~~~ $5282$ ~~~  \\
$K^* \pi$  & ~~~ $4187-i2098$~~~ & ~~~ $4683$~~~  & $-$   & $-$  \\
\hline
\end{tabular}
\label{tab:k1pole}
\end{center}
\end{table}

In Figs. \ref{fig:diak1},  \ref{fig:diakx1}  we  \roca{show the mechanisms for} $\tau^- \to \nu_\tau K^- K_1 (1270)$ decay for the formation of the two  $K_1 (1270)$ states.
Fig. \ref{fig:diak1} shows the decay  $\tau^- \to \nu_\tau K^- K_{1}(1)$  via the process $\tau^- \to \nu_\tau K^{*-} K^{*0}$ (Fig.~\ref{fig:diakx1}(a))
followed by the $K^{*-}$ decay into $K^- \pi^0$  and the merging of the $K^{*0} \pi^0$ into $ K_{1} (1)$.
Fig. \ref{fig:diakx1} shows the decay $\tau^- \to \nu_\tau K^- K_{1} (2)$
via  the process $\tau^- \to \nu_\tau \rho^{-} \rho^0$  followed by the $\rho^{-} $ decay into $K^- K^0$ and the merging of the $ \rho^0 K^0$ into the second  pole of  $K_{1}$, $K_{1} (2)$;
Fig. \ref{fig:diakx1} (b) shows the process $\tau^- \to \nu_\tau \rho^{0} \rho^-$  followed by the $\rho^{0} $ decay into $K^- K^+$ and the merging of the $ \rho^- K^+$ into the second  pole of  $K_{1}$, $K_{1} (2)$.
\

\roca{\subsection{The $VV$ weak production vertex}}

Next, we address  the evaluation of the  $\tau \to \nu_\tau K^{*0} K^{*-} ,\nu_\tau K^{*-} K^{*0}$, and $\tau \to \nu_\tau  \rho^{-} \rho^{0}, \nu_\tau  \rho^{0} \rho^{-}$  amplitudes.
The production is assumed to proceed first from the Cabibbo favored $\bar u d$ production from the $W^-$ which
then hadronizes producing an extra $\bar{q} q$ with the quantum numbers of the vacuum, which are implemented with the $^3P_0$ model \cite{micu,LeYaouanc:1972vsx,bijker}.
In  Ref.~\cite{tdai} the mechanism for hadronization is done in detail.
\roca{Next, we summarize and adapt the formalism to the present case.}
The first step corresponds to the flavor combinations in the hadronization.  \roca{In Ref.~\cite{tdai}} it is shown that $d(\bar{s}s)\bar{u}=(d\bar{s})s\bar{u}$
gives rise to $K^{*0} K^{*-}$,  while $d(\bar{u} u) \bar{u}$ and $d(\bar{d} d) \bar{u}$ give rise to
 $\rho^{-} \rho^{0}$ and $\rho^{0} \rho^{-}$  [see Eq. (4)] of \cite{tdai}).
 The second  step corresponds to the detailed study  of  the spin-angular momentum algebra to combine the quarks for the
 $^3P_0$ $\bar{q}q$ state ($L'=1, S'=1,J'=0$) with a $\bar{d}$  quark in $L=1$  to have finally $s$-wave production of the two mesons.
 In \cite{Dai:2018rra} a similar calculation has been done to discuss the triangle singularity in $\tau^- \to  \nu_\tau \pi^- f_0(980)$ ($a_0(980)$) decays,  but  \roca{with} pseudoscalar-vector production
 from the $W$ rather than two vectors, as we have here.

The elementary quark  \roca{$d\bar u$ production in the tau decay}  is given by
\begin{eqnarray}\label{eq:c}
H= \mathcal{C} L^\mu Q_\mu \,,
\end{eqnarray}
where $\mathcal{C}$ contains the couplings of the weak  interaction  \roca{to be determined later on}.  The  leptonic current is given by
\begin{eqnarray}
L^\mu=\langle {\bar u}_\nu |\gamma^\mu-\gamma^\mu\gamma_5| u_\tau\rangle \,,
\end{eqnarray}
and  the quark current  by
\begin{eqnarray}
Q^\mu=\langle \bar u_d|\gamma^\mu-\gamma^\mu\gamma_5|v_{\bar u}\rangle  \,.
\end{eqnarray}
In the evaluation of \roca{the} decay widths to three final particles, we \roca{find convenient to} evaluate the matrix elements in the frame where the two meson system is at rest, and
we assume that the quark spinors  are at rest  for  the evaluation of the  $Q_\mu$  matrix element  in the same frame \cite{tdai}.
Then we have $\gamma^0 \to 1$,  $\gamma^i \gamma_5 \to \sigma^i$
in terms of bispinors $\chi$, and  after the spin angular momentum combination we  \roca{end up with the following spinor matrix elements:}
\begin{eqnarray} \label{eq:Qu2}
Q_0&=& \langle \chi^{\prime} | 1 | \chi \rangle \roca{\equiv}  M_0  \nonumber \, , \\
Q_i&=&  \langle \chi^{\prime} | \sigma_i |\chi \rangle \roca{\equiv}  N_i \, .
\end{eqnarray}
Denoting for simplicity,
\begin{equation}\label{eq:new}
\overline{L}^{\mu\nu}\roca{\equiv} \overline{\sum} \sum  L^\mu {L^\nu}^\dagger  \, ,
\end{equation}
to obtain the $\tau$ width we  must evaluate
\begin{eqnarray}\label{eq:L}
\overline{\sum} \sum \left|t\right|^2 &=&\overline{\sum} \sum  L^\mu {L^\nu}^\dagger Q_\mu Q_\nu^*  \nonumber \, , \\
&=& \bar{L}^{00}\, M_0~ M^*_0
+\bar{L}^{0i}\,M_0 ~N^*_i
+\bar{L}^{i0} \, N_i ~M^*_0
+\bar{L}^{ij} N_i ~N_j^* \, ,
\end{eqnarray}
with $\overline{L}^{\mu\nu}$ given by
\begin{eqnarray}\label{eq:LL}
\overline{\sum} \sum  L^\mu {L^\nu}^\dagger =\frac{1}{m_{\nu} m_{\tau}}\left( p'^\mu p^\nu  +p'^\nu p^\mu - g^{\mu\nu}p'\cdot p+i \epsilon^{\alpha\mu\beta\nu}p'_\alpha p_\beta\right) \, ,
\end{eqnarray}
where $p,p'$ are the momenta of the $\tau$ and $\nu_\tau$ respectively and  we use the field normalization for fermions of Ref. \cite{mandl}.
\roca{The expression of the amplitudes in terms of the $M_0$ and $N_i$ functions was the main novelty of the work in \cite{tdai}. This formalism has the advantage of filtering the G-parity contributions since the $M_0$ and $N_i$ operators act with defined G-parity as explained below.}

From the work  \cite{tdai}  we obtain  the results for the  $VV, J=1, J'=1$ case,

\begin{equation}\label{eq:oM0}
M_0=\frac{1}{\sqrt{3}}\frac{1}{4\pi} {\cal C}(1 1 1; M,M',M+M')  \, ,
\end{equation}
\begin{eqnarray}\label{eq:oNmu}
N_{\mu} = \frac{1}{\sqrt{6}} \frac{1}{4\pi}  \Big\{ \delta_{M \mu} + 2 \, (-1)^{-\mu-M'} \ \mathcal{C}\left( 1 1 1; M, -\mu, M-\mu \right)
 \mathcal{C}\left( 1 1 1; M', -M-M'+\mu, -M+\mu \right) \Big\}  \, , ~~~~
\end{eqnarray}
where $M, M'$ are the third components of $J, J'$  respectively, and $\mu$ is the index of $N_i$ in spherical basis, with ${\cal C}(\cdots)$ a Clebsch-Gordan coefficient.

In \cite{tdai},  \roca{it was shown} that the order in which the vector  mesons are produced is essential to understand the $G$-parity symmetry  of these reactions,  which is given in Table \ref{tab:sign}.
\begin{table}[h!]
\renewcommand\arraystretch{1.0}
\caption{ Signs resulting in the $M_0$, and $N_{\mu}$ amplitudes for {\color{black} $s$-wave} by permuting the order of the mesons. }
\centering
\begin{tabular}{ccccc }
\toprule[1.0pt]\toprule[1.0pt]
~~~~~~~&~~~~~~~ $PP$~~~~~~~  & ~~~~~~~ $PV$ ~~~~~~~& ~~~~~~~ $VP$~~~~~~~ &~~~~~~~  $VV$~~~~~~~ \\
\hline
$M_0$ & $-$  &  $+$ &  $+$ &  ${\color{black}-}$ \\
\hline
$N_\mu$ &$+$  &  $-$ &  $-$ &  ${\color{black}+}$ \\
\midrule[1.0pt]\midrule[1.0pt]
\end{tabular}\label{tab:sign}
\end{table}
We note from Table \ref{tab:sign}  that $M_0$ changes sign for VV, when exchanging the mesons, while for $N_i$ it is the same. This sign is essential for the conservation of $G$-parity  in the reaction, as we shall see.
On the other hand, at the quark  level  the primary $d\bar{u}$  state produced has $I_3=-1$ and hence $I=1$.  The $G$-parity of a $q\bar{q}$  pair is given by $(-1)^{L+S+I}$. As we mentioned, $L=1, I=1$ and the spin
of the state is $0$ for the $1$ operator and $1$ for the $\sigma^i$ operator  of Eq. \eqref{eq:Qu2}. This means that the term $N_i$  proceeds with $G$-parity  \roca{negative}, while  $M_0$ has $G$-parity
 \roca{positive}.
Since $\pi$, $h_1(1170)$, $h_1(1380)$, $a_1(1260)$, $f_1(1285)$, and $b_1(1235)$ have $G$-parity $-, -, -, -, +, +$ respectively,
then  $\pi^- f_1(1285)$ and $\pi^- b_1(1235)$  proceed with the  \roca{$N_i$} amplitude,   while $\pi^- h_1(1170)$, $\pi^- h_1(1380)$ and $\pi^- a_1(1260)$
will proceed with the  \roca{$M_0$}  term  and there is no simultaneous  contribution of the two terms in these reactions, \roca{implying that the crossed terms in Eq.~\eqref{eq:L} are zero.}
 \roca{W}e shall  see \roca{this} analytically   when evaluating explicitly  the amplitudes for the processes of Figs. \ref{fig:diane1} and \ref{fig:diane2}.

\subsection{ Evaluation of \roca{the} constant $\cal{C}$ }

  \roca{The} global unknown constant $\cal{C}$ in Eq. \eqref{eq:c}  can be determined  from the experimental ratio of $\tau \to \nu_\tau K^{*0}K^{*-}$\roca{, using a similar method as in Ref. \cite{Oset:2018zgc}}.

  In the present work,  the structure of $\overline{\sum} \sum |t|^2$  \roca{for the $\tau$ decay into two vector mesons} is taken from  the results  of  \cite{tdai} for this reaction.
 \roca{If}  we  take the quantization axis along the direction of the neutrino in the $\tau^-$ rest frame  \roca{we} find
\begin{eqnarray}\label{eq:ff2}
\overline{\sum} \sum \left|t\right|^2 &=&  \frac{{\cal{C}}^2}{m_\tau m_\nu} \left(\frac{1}{4\pi}\right)^2
 \left[\left(E_\tau E_\nu + {p}^2 \right){h}^2_i+ \frac{7}{2}\left(E_\tau E_\nu -\frac{1}{3}  {p}^2  \right) \overline{h}^2_i \right]  \,,
 \end{eqnarray}
where   $p$ is the momentum of the $\tau$, or $\nu_\tau$, in the $K^{*0}K^{*-} $ rest frame, given by
\begin{equation}\label{eq:newlabel}
p=p_\nu=p_\tau=\frac{\lambda^{1/2}(m^2_\tau,m^2_\nu,M_{\rm inv}^{2} (K^{*0} K^{*-}))}{2 M_{\rm inv}{(K^{*0} K^{*-})}}\, ,
\end{equation}
and $E_\nu=p$, $E_\tau=\sqrt{m_\tau^2+p^2}$.
\roca{In Eq.~\eqref{eq:ff2}  the coefficients $h_i$  and $\overline{h}_i$ account for the weights of the $VV$ components for $M_0$
and $N_\mu$ respectively and their values are listed in Table \ref{tab:hs}. Note that we are considering only the final s-wave production since, as explained in \cite{tdai}, because of the large vector masses, the expected momenta are very small.}
\begin{table}[h]
\renewcommand\arraystretch{1.1}
\caption{ \roca{The} $h_i$ and $\overline{h}_i$ coefficients for \roca{the} different \roca{$VV$} channels with the two \roca{vectors} in $s$-wave. }
\centering
\begin{tabular}{lcc}
\toprule[1.0pt]\toprule[1.0pt]
~channels~~~&~~~~ $h_i$ (for $M_0$)~~~~  & ~~~~$\overline{h}_i$ (for $N_\mu$) ~~~~ \\
\hline
$K^{*0} K^{*-}$  &   $1$  & $1$ \\
$\rho^{-} \rho^{0}$ &  $\sqrt{2}$   & $0$  \\
\midrule[1.0pt]\midrule[1.0pt]
\end{tabular}
\label{tab:hs}
\end{table}

The mass distribution is given by
\begin{equation}\label{eq:dGdM}
\frac{ d\Gamma}{d M_{\rm inv}{(K^{*0} K^{*-})}} =  \frac{2\,m_\tau 2\, m_\nu}{(2\pi)^3} \frac{1}{4 m^2_\tau}\, p'_\nu {\widetilde p_K}\, \overline{\sum} \sum \left|t\right|^2 \,,
\end{equation}
where ${\widetilde p_K}$ is the momentum of $K^{-}$  in the $K^{*0}K^{*-}$ rest frame  given by
\begin{equation} \label{eq:new2}
\widetilde{p}_K=\frac{\lambda^{1/2}(M_{\rm inv}^{2} (K^{*0} K^{*-}), m_{K^{*0}}^2, m_{K^{*-}}^2)}{2 M_{\rm inv}{(K^{*0} K^{*-})}}\, ,
\end{equation}
and $p'_\nu$  the neutrino momentum in the $\tau$ rest frame
\begin{equation}
p'_\nu=\frac{\lambda^{1/2}(m^2_\tau,m^2_\nu,M_{\rm inv}^{2} (K^{*0} K^{*-}))}{2 m_\tau}\, .
\end{equation}
The experimental branching ratio of the  $\tau \to \nu_\tau K^{*0} K^{*-}$ decay was constructed in Ref. \cite{Oset:2018zgc}  from information in the  PDG \cite{pdg}, with the result,
\begin{eqnarray}\label{eq:Ce}
{\cal B} (\tau \to \nu_\tau K^{*0} K^{*-}) = \frac{1}{\Gamma_\tau} \Gamma(\tau \to \nu_\tau K^{*0} K^{*-}) =(2.1 \pm 0.5)\times 10^{-3},
\end{eqnarray}
from which  we can evaluate the value of the constant ${\cal C}^2$,
\begin{eqnarray}\label{eq:C}
 \frac{{\cal C}^2}{\Gamma_\tau} = \frac{{\cal B} (\tau \to \nu_\tau K^{*0} K^{*-})}{ \int \frac{ d\Gamma}{d M_{\rm inv}{(K^{*0} K^{*-})}}dM_{\rm inv}{(K^{*0} K^{*-})} } = 5.0 \times 10^{-4} ~~[{\rm  MeV}^{-1}]\,  ,
\end{eqnarray}
\roca{Note that the $\tau$ decay into $\nu_\tau K^{*0} K^{*-}$ can only proceed because of the finite width of the $K^*$, since otherwise there would be no available phase space for infinitely narrow $K^*$. Hence it is crucial to fold the width with a realistic spectral function of  the $K^*$ meson, (see Eq.(9) in \cite{Oset:2018zgc}).}
\roca{Note that}, in Ref. \cite{Oset:2018zgc},  the structure of $\overline{\sum} \sum |t|^2$ was assumed to proceed with the dominant term $E_\tau E_\nu -\frac{p^2}{3}$ of Eq. \eqref{eq:ff2} alone,  and hence a somewhat  different ${\cal C}$ constant was obtained.

\subsection{Evaluation of  the triangle diagram}

 In order to evaluate the \roca{triangle} loops of Fig. \ref{fig:diane1},
\roca{we need first the $K^* \to K \pi$ vertex  obtained from the VPP Lagrangian
\begin{equation}
\mathcal{L}_{VPP} = -i g \left < \left[P, \partial_{\mu} P\right]  V^{\mu}\right >  \, ,
\label{eq:vpp}
\end{equation}
with the coupling $g=4.31$ \cite{Oset:2018zgc}, $P$ and $V$ the SU(3) matrices of the pseudoscalar and vector mesons, by means of which we find
\begin{eqnarray}\label{eq:ts1}
 t_{K^{*0} \to \pi^- K^+}=  (2 {\bm k} + {\bm q }) \cdot {\bm \epsilon  } \, g \, ,
\end{eqnarray}
\begin{eqnarray}\label{eq:ts2}
 t_{K^{*-} \to \pi^- K^0}=  -(2 {\bm k} + {\bm q }) \cdot {\bm \epsilon  } \, g \, .
\end{eqnarray}
 We can see that for this vertex we find  a relative minus sign  from Fig. \ref{fig:diane1} (a)  to Fig. \ref{fig:diane1} (b).
 }
We find  convenient to take the $z$ direction  along the momentum $\bm{k}$ of the \roca{final} pion  (see Fig. \ref{fig:mom}). Indeed, in the $\pi A$ rest frame, where we evaluate  the amplitude, ${\bm P}=0$.  The vertex $K^* \to K \pi$ is of the type ${\bm\epsilon}\cdot ({\bm k}+{\bm q}+{\bm k})$
\footnote{ Since in the triangle diagram the  $K^{*0} K^{*-}$ intermediate states  have a small momentum compared to the $K^{*}$ mass \cite{Oset:2018zgc}, we neglect the ${\bm\epsilon}^0$ component,  which was found in \cite{sakairamos} to be an excellent approximation in such a case.}.
On the other hand, the ${\bm q}$ integration  \roca{$ \int  d {\bm q}^i (2\bm  k+\bm  q)\cdots$ of the triangle loop}  \roca{must} necessarily give
something proportional to $ {\bm k}$, which is the only non integrated vector in the loop integral.  Then
 \roca{$ \int  d {\bm q}^i (2\bm  k+\bm  q)\cdots=A \bm{k}$ and contracting with $\bm k$ gives $ \bm k\int  d {\bm q}^i (2+\bm  q\cdot\bm k/\bm k^2)\cdots$.}
Hence, we have an effective vertex of the type ${\bm\epsilon}\cdot {\bm k}$. If the $z$ direction is chosen along ${\bm k}$, this  selects only the ${\bm\epsilon}_z$ component (${\bm\epsilon}_0$ in spherical basis) and  ${\bm\epsilon} \cdot {\bm k}=|{\bm k}|=k$.  This also means that only $M=0$\roca{, for Eqs. \eqref{eq:oM0}, \eqref{eq:oNmu},} contributes in the loop and this allows us to calculate  trivially the $M_0$ and $N_\mu$ amplitudes in that
frame.

\subsubsection{Evaluation of $M_0$ }
For $K^{*0} (M)$ and $K^{*-} (M')$ of Fig. \ref{fig:diane1} (a) and $M=0$, we get from  Eq. \eqref{eq:oM0}
\begin{eqnarray}\label{eq:Q3}
M_0 &\to& \frac{1}{\sqrt{3}}\frac{1}{4\pi} {\cal C}(1 1 1; 0, M',M')     \, .
\end{eqnarray}

On the other hand, in Fig. \ref{fig:diane1} (b) we will have $M'$ of $K^{*-}$  equal to zero and then
\begin{eqnarray}\label{eq:aQ3}
M_0 &\to& \frac{1}{\sqrt{3}}\frac{1}{4\pi} {\cal C}(1 1 1; M, 0, M) =(-1) \frac{1}{\sqrt{3}}\frac{1}{4\pi} {\cal C}(1 1 1; 0, M,  M)   \, .
\end{eqnarray}
We can see that the $M_0$  changes sign  from Fig. \ref{fig:diane1} (a) to Fig. \ref{fig:diane1} (b).  From the sign of Eqs. \eqref{eq:ts1} and \eqref{eq:ts2} and this latter sign,
we can see that the global sign is  the same for these two diagrams.

\roca{Finally, in order to evaluate the final amplitude of the loop diagram we need the vertex $A \to V P$, which  is  of the type \cite{roca05}}
\begin{eqnarray}
g_{A,VP} \, {\bm \epsilon}_V \cdot {\bm \epsilon}_A
\label{eq:gAVPeps}
  \end{eqnarray}
\roca{with   ${\bm \epsilon}_V$ ,${\bm \epsilon}_A$,
 the polarization vector of the vector and axial-vector resonances.
 Note that the couplings of the axials to pseudoscalar and vector are given in \cite{roca05} in terms of well defined G-parity PV sates
(see Tables~\ref{tab:popole}-\ref{tab:k1pole}).
In order to relate those couplings to the charge basis PV states that we are using, w}e need to write our states for the axial vector mesons
in terms of their vector pseudoscalar components forming states of  \roca{well defined} $C$ and $G$ parity:
\begin{eqnarray}\label{eq:Gp1}
 |I=0,C=-\rangle &=& \sqrt{\frac{1}{2}} |\bar{K}^*\roca{K}  + K^*\bar{K}\rangle = -\frac{1}{2}(K^{*+} K^- +  K^{*0} \bar{K}^0 + K^{*-} K^+ +  \bar{K}^{*0} K^0 )   \nn \,  \\
 |I=0,C=+\rangle &=& \sqrt{\frac{1}{2}} |\bar{K}^* K - K^*\bar{K}\rangle = -\frac{1}{2}(K^{*+} K^- +  K^{*0} \bar{K}^0 - K^{*-} K^+  - \bar{K}^{*0} K^0 )   \, ,~~
 \end{eqnarray}
\begin{eqnarray}\label{eq:Gp2}
 |I=1,C=-\rangle &=& \sqrt{\frac{1}{2}} |\bar{K}^* K + K^*\bar{K}\rangle = -\frac{1}{2}(K^{*+} K^- -  K^{*0} \bar{K}^0 + K^{*-} K^+  -  \bar{K}^{*0} K^0 )   \nn \,  \\
 |I=1,C=+\rangle &=& \sqrt{\frac{1}{2}} |\bar{K}^* K - K^*\bar{K}\rangle = -\frac{1}{2}(K^{*+} K^- -  K^{*0} \bar{K}^0 - K^{*-} K^+  + \bar{K}^{*0} K^0 )   \, ,~~
 \end{eqnarray}
Since  $G=(-1)^I C$, we  \roca{need} from the sum of Figs. \ref{fig:diane1} (a) and \ref{fig:diane1} (b), \roca{the following combinations}.
\begin{eqnarray} \label{eq:Gp3}
 I=0, C=-, G=-  \, ,& \quad  g_{A,K^{*-}K^+}+g_{A,K^{*0} \bar{K}^0} = -g_{A, K^{*}\bar{K}} \nn \, , \\
 I=0,C=+, G=+    \, ,& \quad  g_{A,K^{*-}K^+}+g_{A,K^{*0} \bar{K}^0} = 0  \nn \, , \\
 I=1,C=-, G=+    \, ,& \quad  g_{A,K^{*-}K^+}+g_{A,K^{*0} \bar{K}^0} = 0 \nn \, , \\
 I=1,C=+, G=-    \, ,& \quad  g_{A,K^{*-}K^+}+g_{A,K^{*0} \bar{K}^0} = g_{A, K^{*}\bar{K}} \, ,
  \end{eqnarray}
\roca{Note} that Eq.~\eqref{eq:gAVPeps} implies that the $M'$ third component of Eq. \eqref{eq:Q3} becomes the $M_A$ third component of the axial  vector $A$.
Then the $M_0$ contribution to $h_1$ and $a_1$ from the  $K^* \bar{K}^* $ loop becomes
\begin{eqnarray} \label{eq:tm0}
t_{M_0}= {\cal C}  \,g \, k \, h_{K^{*0} K^{*-}}\, {\cal C}(1 1 1; 0, M_A, M_A)   {\color{blue} (\pm 1)}  \,g_{A, K^{*}\bar{K}}  \,  t_L (K^* \bar{K}^*)   \, ,
\end{eqnarray}
with the $+,-$ sign for $a_1$ and $h_1$  production, respectively,  \roca{and}
\begin{equation}
\begin{split}
\label{eq:int2d3}
t_L&=\int\frac{d^3q}{(2\pi)^3}\,\frac{2+\bm{k}\cdot\bm{q}/|\bm{k}\,|^2}{8\,\omega_1\omega_2\omega_3}\,\frac{1}{k^0-\omega_3-\omega_1+i\epsilon}\,\frac{1}{P^0-\omega_1-\omega_2+i\epsilon}\\
&\times \frac{2P^0\omega_2+2k^0\omega_3-2(\omega_2+\omega_3)(\omega_2+\omega_3+\omega_1)}{(P^0-\omega_2-\omega_3-k^0+i\epsilon)(P^0+\omega_2+\omega_3-k^0-i\epsilon)}\ ,
\end{split}
\end{equation}
\roca{is the triangle loop function}
which appears after an analytical calculation of the $q^0$ integral \cite{acetidias}. For the case of the diagrams in Fig.\ref{fig:diane1} (a) the states $1, 2, 3$ correspond to $K^{*0}$, $K^{*-} $, $K^{+}$, respectively. \roca{In Eq.~\eqref{eq:int2d3}}  $\omega_1=\sqrt{\bm{q}^2+m_{1}^2}$, $\omega_2=\sqrt{\bm{q}^2+m_2^2}$ and $\omega_3=\sqrt{(\bm{q}+\bm{k})^{2}+m_3^2}$ are the energies of the $1$, $2$  and $3$  states in the loop respectively,
$P^0=M_{\rm inv}(\pi^- A)$, and
\begin{equation}
k^0=\frac{M^2_{\rm inv}(\pi^- A)+m_{\pi}^2-m^2_A}{2 M_{\rm inv}(\pi^- A)},
\end{equation}
\begin{equation} \label{eq:25}
k=\frac{\lambda^{1/2}(M^2_{\rm inv}(\pi^- A), m_{\pi}^2, m^2_A)}{2 M_{\rm inv}(\pi^- A)}.
\end{equation}
We also account for the $K^{*0}$, $K^{*-}$ widths by replacing $\omega_1\to\omega_1+i\Gamma_{K^*}/2$, $\omega_2\to\omega_2+i\Gamma_{K^*}/2$
in the propagators involving  $\omega_{K^{*}}$ in the actual calculation.
Similarly, we can get the  triangle amplitude for Fig.\ref{fig:diane1} (b) case.

\subsubsection{Evaluation of $N_i$ }
In spherical basis,  $N_i \to N_\mu~(\mu=0,\pm 1)$,  is given by Eq. \eqref{eq:oNmu}. Once again we chose now  ${\bm k}$ in the $z$ direction and thus force $M=0$.
On the other hand, as we did before, $M'$ becomes $M_A$,
the axial  vector polarization, since the $s$-wave coupling of $A \to P V~ ({\bm \epsilon}_A \cdot {\bm \epsilon}) $ implies the same $A$ and $V$ polarization. Then Eq. \eqref{eq:oNmu}  becomes
\begin{eqnarray}\label{eq:Nmu}
N_{\mu} = \frac{1}{\sqrt{6}} \frac{1}{4\pi}  \Big\{ \delta_{\mu 0} + 2 \, (-1)^{-\mu-M_A} \ \mathcal{C}\left( 1 1 1; 0, -\mu, -\mu \right)
 \mathcal{C}\left( 1 1 1; M_A, -M_A+\mu, \mu \right) \Big\}  \, . ~~~~
\end{eqnarray}
Contrary to what happens with the $M_0$ component,  we can see that  $N_\mu$  does not change   sign  when we exchange $K^{*0}K^{*-} \to K^{*-}K^{*0}$ in the loop of Fig. \ref{fig:diane1} (a) and Fig. \ref{fig:diane1} (b). Then, considering the different sign  in the vertex $K^* \to K \pi$  of  Eqs. \eqref{eq:ts1} and \eqref{eq:ts2},
we get the combination $g_{A,K^{*-}K^+} -g_{A,K^{*0} \bar{K}^0}$,   which in view of Eqs. \eqref{eq:Gp1} and  \eqref{eq:Gp2}
provides,
\begin{eqnarray}
 I=0, C=-, G=-  \, ,& \quad  g_{A,K^{*-}K^+} -g_{A,K^{*0} \bar{K}^0} = 0 \nn \, , \\
 I=0,C=+, G=+    \, ,& \quad  g_{A,K^{*-}K^+} -g_{A,K^{*0} \bar{K}^0} = g_{A, K^{*}\bar{K}}  \nn \, , \\
 I=1,C=-, G=+    \, ,& \quad  g_{A,K^{*-}K^+} -g_{A,K^{*0} \bar{K}^0} = -g_{A, K^{*}\bar{K}} \nn \, . \\
 I=1,C=+, G=-    \, ,& \quad  g_{A,K^{*-}K^+} -g_{A,K^{*0} \bar{K}^0} = 0 \, .
\end{eqnarray}
This  shows explicitly  that with $G$-parity positive \roca{axials} only the $N_i$  term  contributes, as we saw at the beginning  at the quark level,  while for $G$ negative \roca{axials} only
the $M_0$ term contributes.

From  Eq. \eqref{eq:Nmu} we can calculate the $N_i$ components in cartesian basis and we find

 \begin{eqnarray}\label{eq:cn1}
& N_1=\frac{1}{\sqrt{2}} (N_{-1}-N_{+1}) = \frac{1}{\sqrt{2}} &
      \left\{
    \begin{array}{ll}
     -\frac{1}{\sqrt{6}} \frac{1}{4\pi} \, , &~~~~~~~~ (M_A=1)\, \\[2mm]
     0 \, ,&~~~~~~~~ (M_A=0)\, \\[2mm]
   \frac{1}{\sqrt{6}} \frac{1}{4\pi} \, , &~~~~~~~~(M_A=-1)\,\\[2mm]
    \end{array}
   \right.  \,
 \end{eqnarray}
 \begin{eqnarray}\label{eq:cn2}
& N_2=\frac{i}{\sqrt{2}} (N_{-1}+N_{+1}) = \frac{1}{\sqrt{2}} &
      \left\{
    \begin{array}{ll}
     \frac{1}{\sqrt{6}} \frac{1}{4\pi} \, , &~~~~~~~~ (M_A=1)\, \\[2mm]
     \frac{2}{\sqrt{6}} \frac{1}{4\pi} \, ,&~~~~~~~~ (M_A=0)\, \\[2mm]
   \frac{1}{\sqrt{6}} \frac{1}{4\pi} \, , &~~~~~~~~(M_A=-1)\,\\[2mm]
    \end{array}
   \right.  \,
 \end{eqnarray}
\begin{eqnarray}\label{eq:cn3}
      \begin{array}{ll}
N_3=N_0=\frac{1}{\sqrt{6}} \frac{1}{4\pi}\, , & ~~~~~~~~ ( {\rm for ~~any}~~ M_A)\, \\[2mm]
    \end{array}
   \end{eqnarray}

\subsection{Incorporation of intermediate $\rho\rho$ states }

\roca{For the production of negative G-parity axial vector mesons, we  must also consider the  $\rho^0 \rho^-$ diagrams of Fig.~\ref{fig:diane2} in addition to Fig.~\ref{fig:diane1}}.
  For this  we need  the $h_i$ coefficients of table \ref{tab:hs}. Recall that in this case
only the $M_0$  term contributes. Next we need the $\rho^- \to \pi^- \pi^0$,  $\rho^0 \to \pi^- \pi^+$ vertices \roca{obtained} from the Lagrangian in Eq. \eqref{eq:vpp},
\begin{eqnarray}\label{eq:ats1}
 t_{\rho^- \to \pi^- \pi^0}= \sqrt{2} \, g \, (2 {\bm k} + {\bm q }) \cdot {\bm \epsilon  }  \, ,
\end{eqnarray}
\begin{eqnarray}\label{eq:ats2}
 t_{\rho^0 \to \pi^- \pi^+}=  -\sqrt{2} \, g \,  (2 {\bm k} + {\bm q }) \cdot {\bm \epsilon  }  \, .
\end{eqnarray}
Since $M_0$ changes sign from $\rho^- \rho^0$  to $\rho^0 \rho^-$ production \roca{(see Table \ref{tab:sign})}, this sign and the  relative one of Eq. \eqref{eq:ats1},  \eqref{eq:ats2}  cancel and we get the factor in the sum of the loops
\begin{eqnarray}
g_{A,\rho^{0} \pi^0}+g_{A,\rho^{-} \pi^+}  \,.
\end{eqnarray}
To relate  \roca{these couplings in charge basis} to the coupling of $A$ to $\rho \pi$  in isospin basis, evaluated in \cite{roca05}, we recall that the isospin multiplets are $(-\pi^+, \pi^0, \pi^-)$,   $(-\rho^+, \rho^0, \rho^-)$. Then we have
\begin{eqnarray} \label{eq:rho}
 | \rho \rho, I=0, I_3=0 \rangle = -\frac{1}{\sqrt{3}} (\rho^+ \pi^-  + \rho^- \pi^+ + \rho^0 \pi^0)  \nn \, , \\
 | \rho \rho, I=1, I_3=0 \rangle = \frac{1}{\sqrt{2}} (\rho^- \pi^+  - \rho^+ \pi^-)   \, .
 \end{eqnarray}
Then
 \begin{eqnarray}\label{eq:trho}
    \begin{array}{ll}
     g_{A,\rho^{0} \pi^0} + g_{A,\rho^{-} \pi^+}= -\frac{2}{\sqrt{3}} g_{A,\rho \pi}  \, , &~~~~~~~~ {\rm for} ~~I=0~~ (h_1) \, \\[2mm]
     g_{A,\rho^{0} \pi^0} + g_{A,\rho^{-} \pi^+}= \frac{1}{\sqrt{2}} g_{A,\rho \pi}  \, , &~~~~~~~~ {\rm for} ~~ I=1 ~~(a_1) \, \\[2mm]
    \end{array} \, .
 \end{eqnarray}
The $\rho\pi$ channel only contributes to these two states that have negative $G$-parity.

Thus,  \roca{in order to account for} the coherent sum of $K^{*-}K^{*0}$ and $\rho^{-}\rho^{0}$ we \roca{can use} $t_{M_0}$ of Eq. \eqref{eq:tm0}  \roca{but performing the following substitution},
\begin{eqnarray}
 g_{A, K^{*}\bar{K}} \,  t_L(K^* \bar{K}^*)  \to  g_{A, K^{*}\bar{K}} \,  t_L(K^* \bar{K}^*) - 2 \,D \,g_{A,\rho \pi} t_L (\rho \rho) \, ,
\end{eqnarray}
with
 \begin{eqnarray}
 D=
   \left\{
      \begin{array}{ll}
     -\frac{2}{\sqrt{3}}  \, , &~~~~~~~~ {\rm for} ~~I=0~~ (h_1) \, \\[2mm]
     \frac{1}{\sqrt{2}}  \, , &~~~~~~~~ {\rm for} ~~ I=1 ~~(a_1) \,
    \end{array}
    \right.  \, .
 \end{eqnarray}

Next we must perform the sum of Eq. \eqref{eq:L}  independently,   $\bar{L}^{00}\, M_0~ M^*_0$  for negative $G$-parity $A$ states and $\bar{L}^{ij} N_i ~N_j^*$ for positive $G$-parity $A$ states.
By using Eq. \eqref{eq:LL} and  Eqs. \eqref{eq:Q3}, \eqref{eq:cn1}, \eqref{eq:cn2}, \eqref{eq:cn3} and summing over the  $M_A$ components we obtain:
\begin{itemize}
  \item [a)] $G$-parity positive \roca{axial} states:
\begin{eqnarray}\label{eq:po}
\overline{\sum} \sum \left|t\right|^2 &=&  \frac{{\cal{C}}^2}{m_\tau m_\nu} \,\frac{1}{(4\pi)^2} \, \frac{7}{6} \, \left(E_\tau E_\nu -\frac{1}{3}  {p}^2  \right) \,   g^2 \,  k^2 \,  |g_{A, K^* \bar{K}}|^2 \,  |t_L(K^* \bar{K}^* )|^2   \, ,
 \end{eqnarray}
 \item [b)] $G$-parity negative  \roca{axial} states:
 \begin{eqnarray}\label{eq:ne}
\overline{\sum} \sum \left|t\right|^2 &=&  \frac{{\cal{C}}^2}{m_\tau m_\nu} \,\frac{1}{(4\pi)^2} \, \frac{1}{3} \, (E_\tau E_\nu + {p}^2) \, g^2 \,  k^2 \,  \nn \\
&\times&  |g_{A, K^* \bar{K}} t_L(K^* \bar{K}^*)- 2 \, D g_{A, \rho \pi} t_L(\rho\rho )|^2   \, ,
 \end{eqnarray}
\end{itemize}

We should note that the $\epsilon^{\alpha\mu\beta\nu}p'_\alpha p_\beta$  term of Eq. \eqref{eq:LL} does not contribute in $M_0$ since $\mu=0$ and $ p'_\alpha p_\beta (p_\tau p_\nu)$  will be spatial
and ${\bm p}_\tau {\bm p}_\nu$ are the same in the frame we work. For  $N_i$,  $\alpha$ or $\beta$ must be zero and we have just one vector ${\bm p}_\nu$ that cancels  in the phase space integration.
For the same reason,  the term  with $p_{\nu i}p_{\nu j}$  becomes $\frac{1}{3} {\bm p}^2_\nu \delta_{ij}$ upon  integration over phase space.


\subsection{$\tau^- \to  \nu_\tau K^- K_1(1270)$}
In \cite{roca05}  two states corresponding to $K_1$ were found \roca{and the pole positions were refined in \cite{geng07}}, one of them at $1195$ MeV coupling  mostly to $K^* \pi$, and another one at $1284$ MeV coupling
 mostly to $\rho K$.  \roca{Proceeding analogously to the previous cases,} the terms  that go like $L^{0i}$, of $\epsilon^{\alpha\mu\beta\nu}$ cancel again in the integration over phase space, and  we obtain:
\begin{itemize}
 \item [a)] $K_{1} (1) $ state:
\begin{eqnarray}\label{eq:k1a}
\overline{\sum} \sum \left|t\right|^2 &=&  \frac{{\cal{C}}^2}{m_\tau m_\nu} \frac{1}{(4\pi)^2} \,  g^2 \,  k^2 \, |t_L(K^* \bar{K}^* )|^2  \, \frac{1}{3} \,   |g_{K_1, K^*\pi}|^2  \,  \frac{1}{2}             \nn \\
 &\times & \left[\frac{1}{3}\left(E_\tau E_\nu + {p}^2 \right) + \frac{7}{6} \left(E_\tau E_\nu -\frac{1}{3}  {p}^2  \right)   \right]    \, ,
 \end{eqnarray}
  \item [b)]  $K_{1} (2) $ state:
  \begin{eqnarray}\label{eq:k1b}
\overline{\sum} \sum \left|t\right|^2 &=&  \frac{{\cal{C}}^2}{m_\tau m_\nu} \frac{1}{(4\pi)^2} \,  g^2 \,  k^2 \, |t_L(\rho\rho )|^2  \, (\sqrt{2})^2  \frac{1}{3}\left(E_\tau E_\nu + {p}^2 \right)   \, \frac{4}{3} \,   |g_{K_1, \rho K}|^2     \, ,
 \end{eqnarray}
\end{itemize}

For  $\tau^- \to \nu_\tau \pi^- A$ decay,  the   differential  mass distribution for  $M_{\rm inv}(\pi^- A)$  is given by
\begin{eqnarray}\label{eq:dG1}
\frac{1}{\Gamma_{\tau}} \frac{d \Gamma}{d M_{\text{inv}}(\pi^- A)} =  \frac{1}{\Gamma_{\tau}} \, \frac{1}{(2 \pi)^3} \, \frac{2 m_\tau 2 m_\nu}{4 m^2_\tau}  p_{\nu}\, \widetilde{p}_{\pi}
\overline{\sum} \sum \left|t\right|^2 \, ,
\end{eqnarray}
with
\begin{eqnarray}
p_{\nu} = \frac{\lambda^{1/2} (m_{\tau}^2,m^2_\nu, M^2_{\text{inv}}(\pi^- A))}{2 m_{\tau} },	\quad
\widetilde{p}_{\pi}=\frac{\lambda^{1/2} ( M^2_{\text{inv}}(\pi^- A),m^2_{\pi},m^2_{A})}{2 M_{\text{inv}}(\pi^- A)}   \,.
\end{eqnarray}
and $p$ is the momentum of the $\tau$, or $\nu_\tau$, in the $ \pi^- A$ rest frame, given by
\begin{eqnarray}
\widetilde{p}=\frac{\lambda^{1/2} (m_{\tau}^2,m^2_\nu, M^2_{\text{inv}}(\pi^- A))}{2 M_{\text{inv}}(\pi^- A)}   \,.
\end{eqnarray}

Similarly, for the $\tau^- \to \nu_\tau K^- K_1(1270)$ decay,  we can get the   differential  mass distribution for $M_{\rm inv}(K^- K_1(1270))$.

Note that the term ${m_\tau m_\nu}$ in the numerator of Eq. \eqref{eq:dG1} cancels the same factor in the denominator of Eqs. \eqref{eq:po}, \eqref{eq:ne}, \eqref{eq:k1a}, and \eqref{eq:k1b}.
In Eq. \eqref{eq:dG1} we have the \roca{same} factor $\frac{{\cal C}^2}{\Gamma_{\tau}}$ \roca{from Eq.\eqref{eq:C}}  and thus we can provide absolute values for the mass
distributions.

\section{Results}

\begin{figure}[ht!]
\includegraphics[scale=0.9]{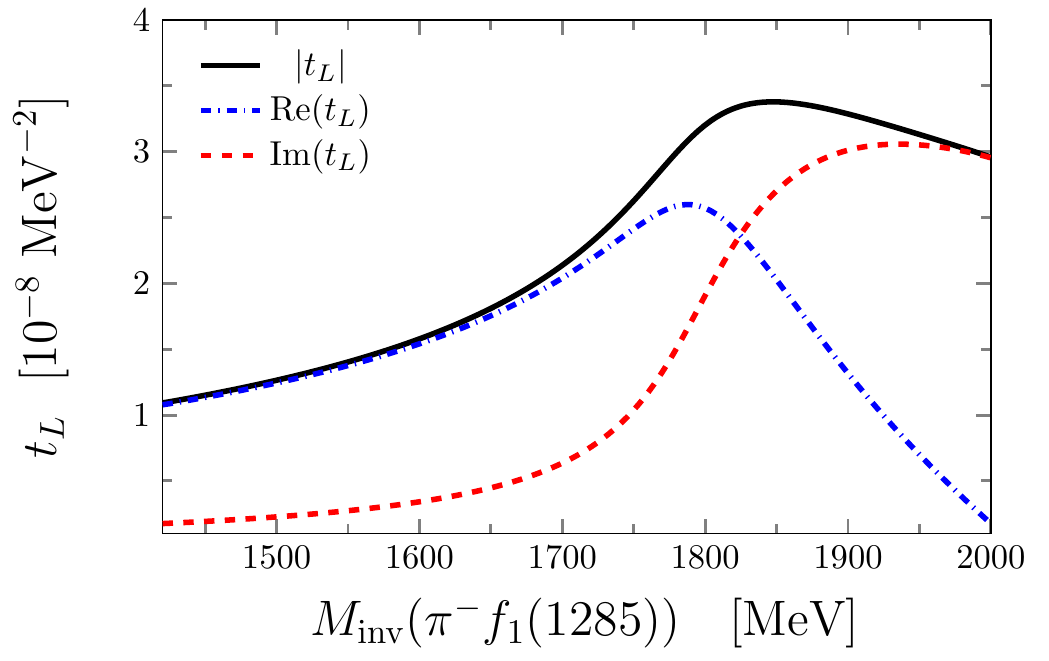}\\
\caption{Triangle amplitude  ${\rm Re} (t_L)$, ${\rm Im}(t_L)$ and $|t_L|$ for  $\tau^-\to\nu_\tau \pi f_1(1285)$ decay, taking $M_{\rm inv}(f_1)$=1229.5 MeV     }
\label{fig:tL}
\end{figure}

\begin{figure}[ht!]
\includegraphics[width=.7\linewidth]{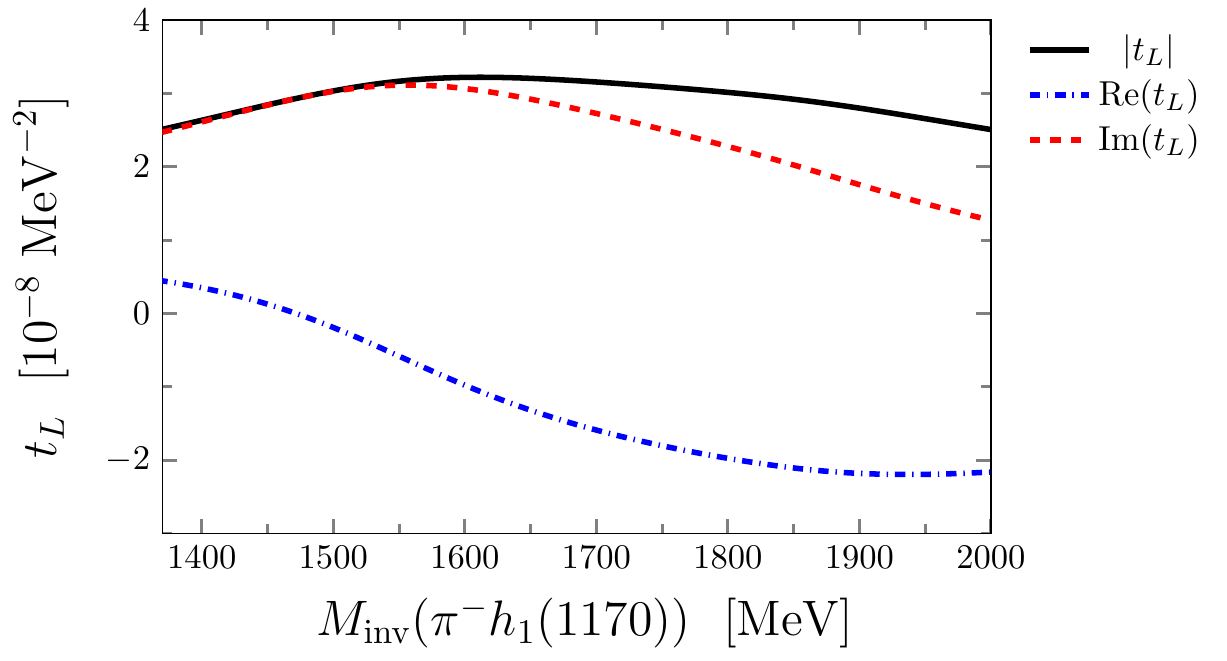}\\
\caption{Same as Fig.~\ref{fig:tL} for $\tau^-\to\nu_\tau \pi h_1(1170)$ decay.}
\label{fig:tLh1}
\end{figure}

\roca{
First we show in Fig.~\ref{fig:tL} the triangle loop  in Eq.~\eqref{eq:int2d3}, for the $\tau^-\to\nu_\tau \pi^- f_1(1285)$ case, as a function of the $\pi^-f_1$ invariant mass, $M_{\rm inv}(\pi^-f_1)$.
Note that there is a large increase of the strength at around the region of interest at the present work, ($M_{\rm inv}(\pi^-f_1)=m_\tau=1777$MeV, which will push the invariant mass distributions for the decays considered in the present work to the higher energy region of the spectrum, as we will see below. As already discussed in \cite{Oset:2018zgc} the origin of this increase is twofold: first because of the presence of a nearby triangular singularity and, second, because of the presence of the $K^* \bar K^*$ threshold. Both effects are implicitly properly taken into account in the  evaluation of the triangle loop in the present work.  In Fig.~\ref{fig:tLh1} we see the triangle loop for $\rho\rho\pi$ as internal lines, for the $h_1(1170)$ in the final state. In this case we see that the enhancement is smaller because there is no a nearby singularity but some strength is visible from the $\rho\rho$ threshold.}

\begin{figure}[ht!]
\includegraphics[scale=1.]{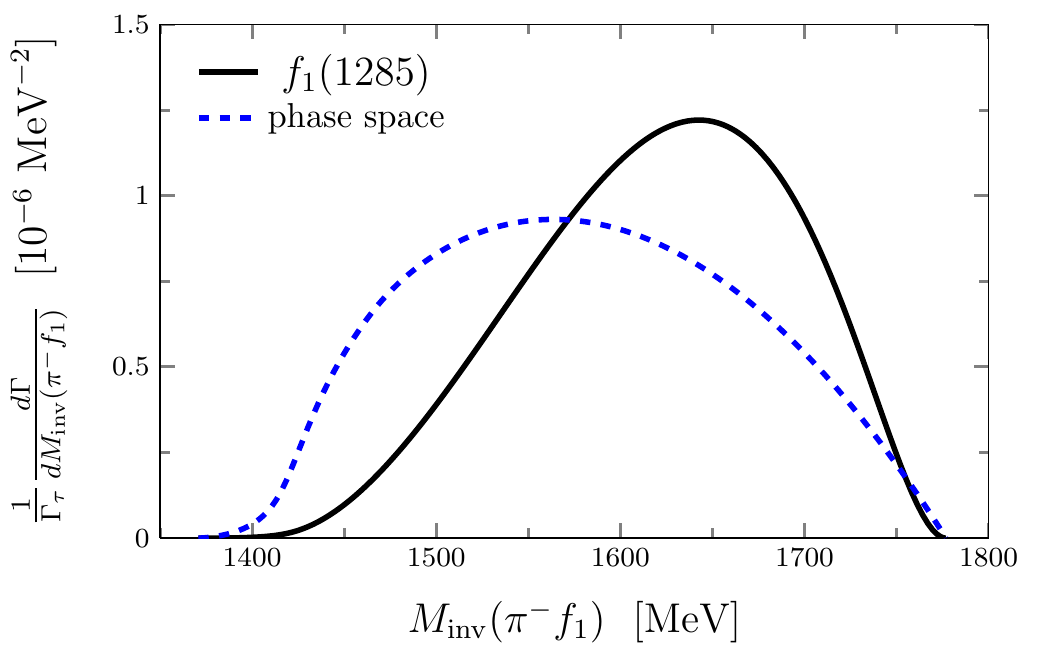}
\caption{Mass distribution for $\tau^- \to \nu_\tau \pi^- f_1(1285)$ decay  }
\label{fig:masf1}
\end{figure}

\begin{figure}[ht!]
\includegraphics[scale=1.]{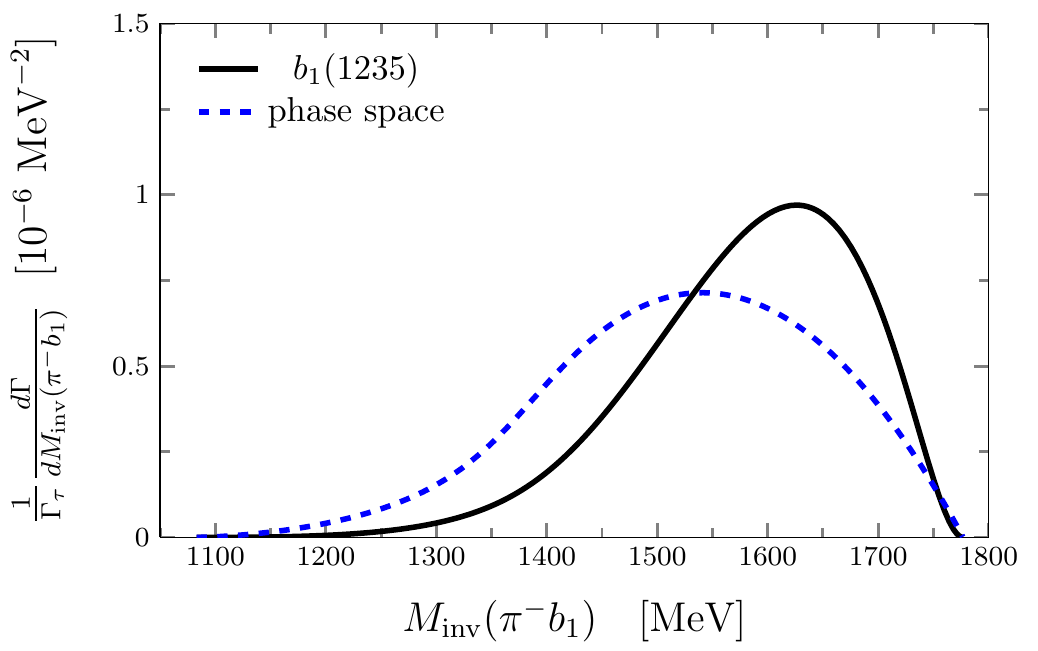}
\caption{Mass distribution for $\tau^- \to \nu_\tau \pi^- b_1(1235)$  decay }
\label{fig:masb1}
\end{figure}

\begin{figure}[ht!]
\includegraphics[scale=1.]{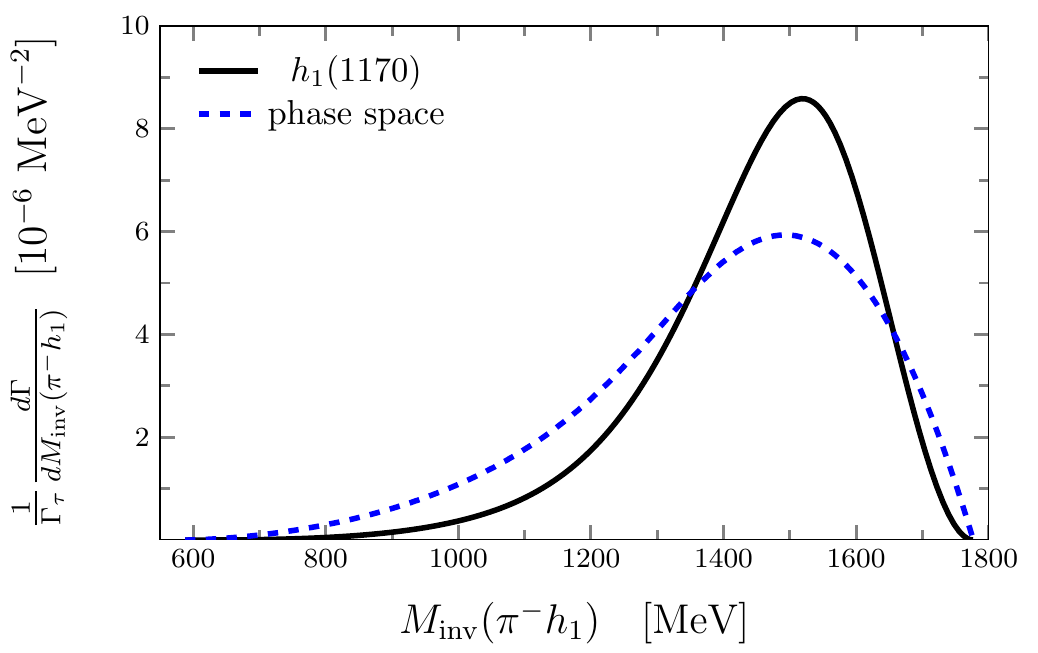}
\caption{Mass distribution for $\tau^- \to \nu_\tau \pi^- h_1(1170)$  decay }
\label{fig:mash1}
\end{figure}

\begin{figure}[ht!]
\includegraphics[scale=1.]{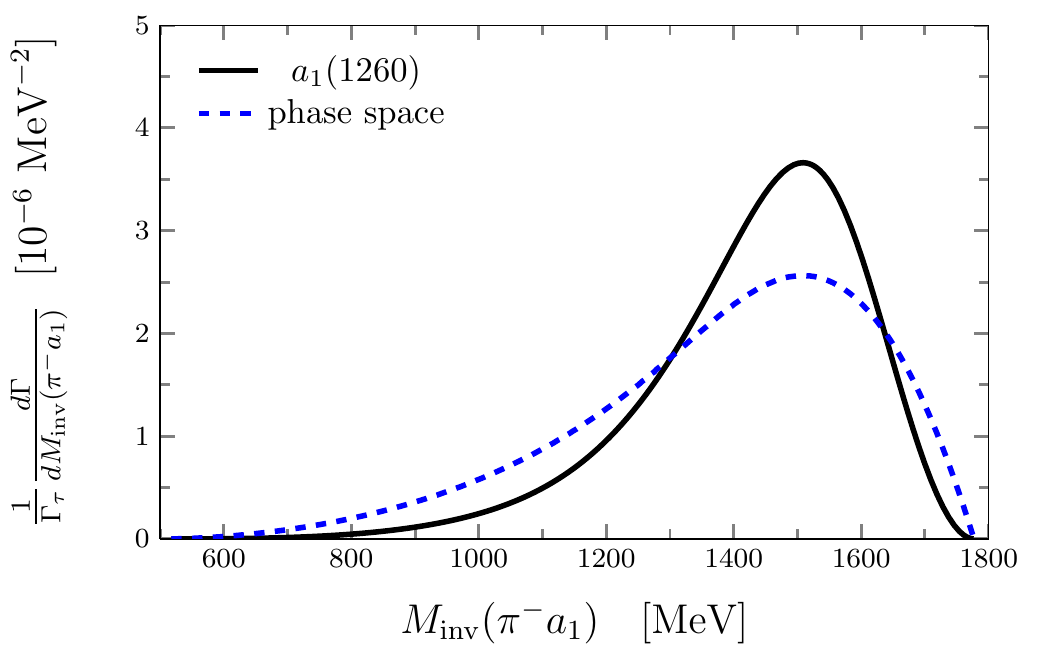}
\caption{The mass distribution for $\tau^- \to \nu_\tau \pi^- a_1(1260)$  decay }
\label{fig:masa1}
\end{figure}

\begin{figure}[ht!]
\includegraphics[scale=1.]{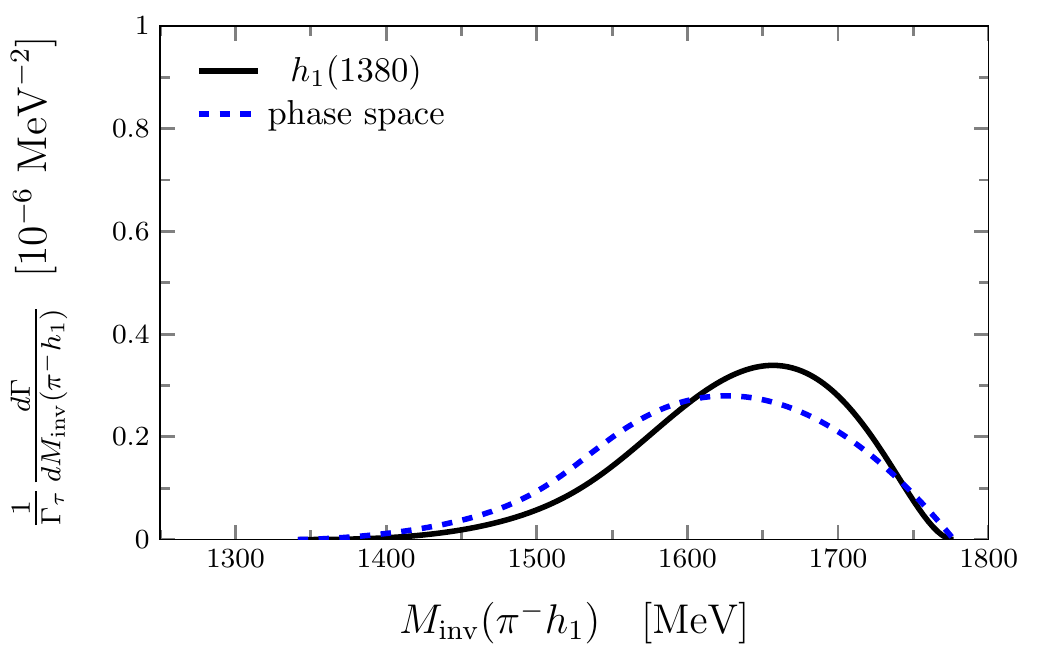}
\caption{The mass distribution for $\tau^- \to \nu_\tau \pi^- h_1(1380)$  decay }
\label{fig:mash1b}
\end{figure}

\begin{figure}[ht!]
\includegraphics[scale=1.]{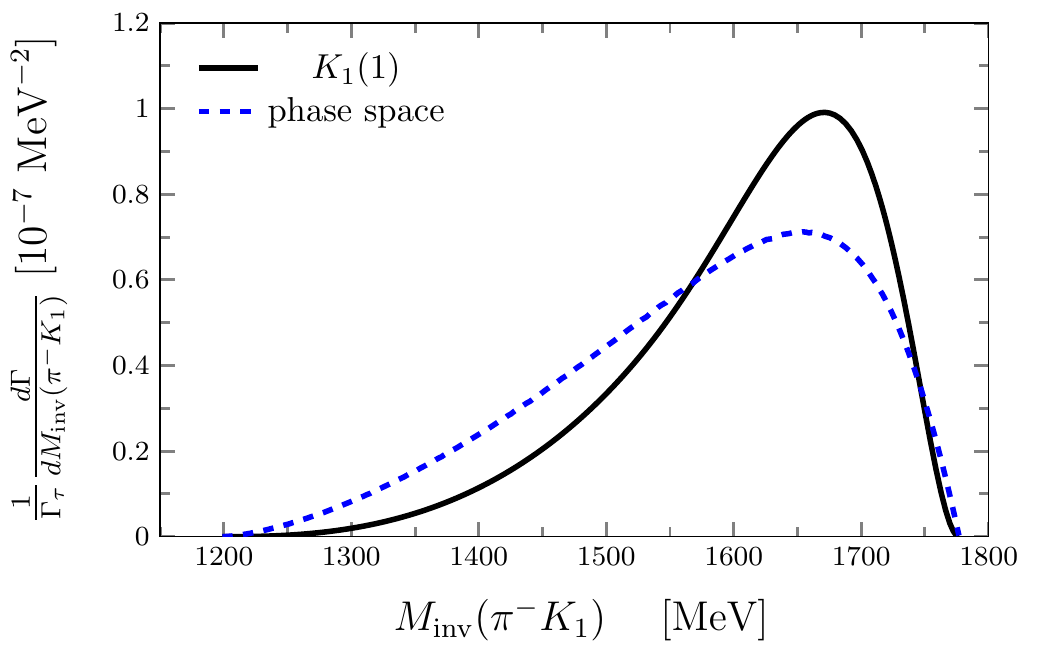}
\caption{The mass distribution for $\tau^- \to \nu_\tau \pi^- K_1(1)$  decay }
\label{fig:mask1}
\end{figure}

\roca{
In Figs.~\ref{fig:masf1}-\ref{fig:mask1}  we show the pseudoscalar-axial, $P A$, invariant mass distributions
of the different
$\tau^- \to  \nu_\tau P A$ decays.
We have considered the finite width of the axial resonances by folding the  invariant mass distributions with the corresponding axial meson spectral function:
 \begin{eqnarray}\label{eq:sc1}
\frac{d\Gamma_{\tau\to \nu_\tau P A}}{dM_{\rm inv} (P A)}&=&\frac{1}{N} \int_{(M_A-2\Gamma_A)^2}^{(M_A + 2\Gamma_A)^2} dm^2\, {\rm Im}  D(m)
\frac{d\Gamma(m)}{dM_{\rm inv} (P A)} \,,
 \end{eqnarray}
where $D(m)$ is the axial-vector propagator,
\begin{eqnarray}
D(m)=\frac{1}{m^2-M_A^2+ i \,\Gamma_A  \, m_A}  \,,\quad  N= \int_{(M_A-2\Gamma_A)^2}^{(M_A + 2\Gamma_A)^2} dm^2 {\rm Im}  D(m) \,,
 \end{eqnarray}
This folding is particularly relevant for the decays into $K_1$ because of the little and null available phase space for the $K_1(1)$ and
$K_1(2)$ respectively. Actually $K_1(2)$ can only proceed because of its tail. }

\begin{figure}[ht!]
\includegraphics[scale=1.]{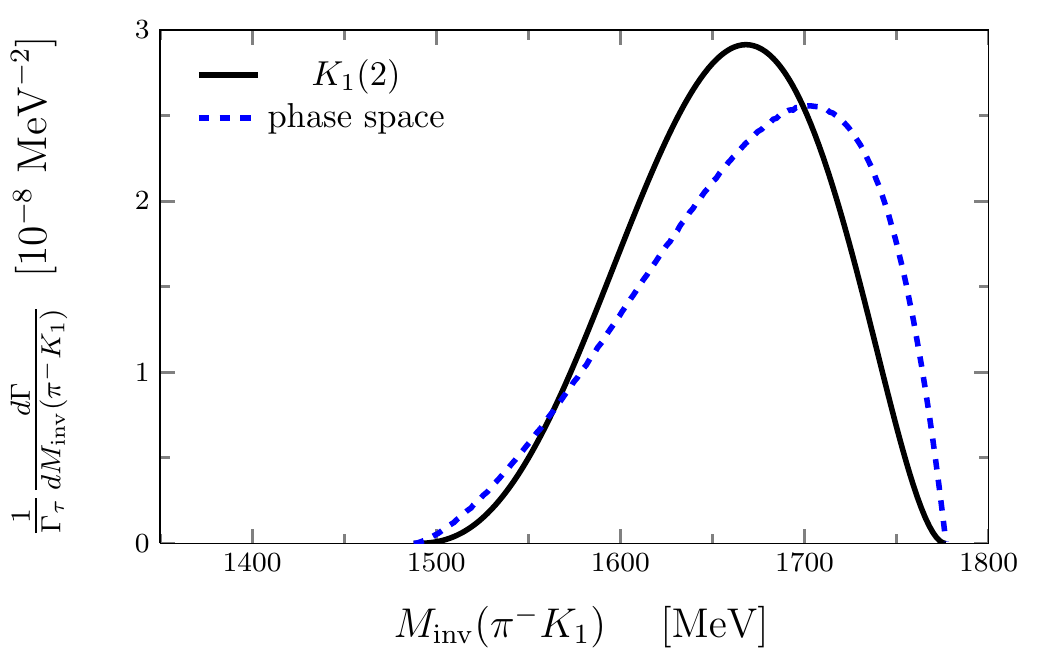}
\caption{The mass distribution for $\tau^- \to \nu_\tau \pi^- K_1(2)$  decay }
\label{fig:mask1b}
\end{figure}
\roca{
 In  Figs.~\ref{fig:masf1}-\ref{fig:mask1} we also plot  (dashed  line)  the  phase  space distribution normalized to  the  area below the  full calculation.
We can see that the strengths of the full calculations are  moved to the higher mass of the spectrum, as a consequence of the particular increase of strength at these energies of the triangle loop function.}
Fig.~\ref{fig:mask1b} for $K_1(2)$ production is an exception tied to the fact that only the tail of the resonance allows its production.

\roca{  In Table \ref{tab:ratio} we show the integrated branching ratios for the different $\tau\to \nu_\tau P A$ decay channels.}
\begin{table}[h!]
\renewcommand\arraystretch{1.0}
\caption{The branching ratios for  $\tau \to \pi^- A, K^- K_1$ decays }
\centering
\begin{tabular}{cc}
   \hline  \hline
  & ${\cal BR}$  \\
       \hline
  ~~~~~~~~$h_1(1170)$~~~~~~~~ & ~~~~~~~~$ 3.0 \times 10^{-3}$ ~~~~~~~~ \\
        $a_1(1260)$ & $ 1.36 \times 10^{-3}$ \\
       $b_1(1235)$ &$ 2.39 \times 10^{-4}$  \\
         $f_1(1285)$ & $ 2.37 \times 10^{-4}$  \\
  $h_1(1380)$ & $ 5.88  \times 10^{-5}$  \\
   $K_1 (1)$  & $ 2.07  \times 10^{-5}$  \\
 $K_1 (2)$  & $ 4.11  \times 10^{-6}$  \\
  \hline  \hline
\end{tabular}
\label{tab:ratio}
\end{table}
\roca{In Ref.~\cite{Oset:2018zgc} a careful error analysis was performed for the $f_1(1285)$ and an error of about 40\% was obtained. For the present calculation, since the sources of uncertainty are similar to those of Ref.~\cite{Oset:2018zgc}, we can also expect an error of the order  of 40\%
to the values shown in Table~\ref{tab:ratio}. Of the branching ratios calculated in the present work only the one for $\tau^- \to  \nu_\tau \pi^- f_1(1285)$ has been experimentally measured \cite{pdg} giving $(3.8\pm1.4)\times10^{-4}$, which compares well with the value we obtain for that channel within uncertainties.
For the channels not yet measured, even though the branching ratios obtained for some of them seem small, they are of the same order as many of the already experimentally measured hadronic decays reported by the PDG \cite{pdg}.}

\roca{
The mass distributions and the branching ratios of table \ref{tab:ratio} are non-trivial and genuine predictions because, first, they crucially depend on the axial-vector resonance couplings to VP which are a non-trivial output of the chiral unitary model \cite{roca05} and consequence of the dynamical origin of these resonances. And, second, because of the non-trivial shape of the triangular mechanism and the enhancement due to nearby singularities when present. Therefore, experimental measurements of these decays could check the dynamical origin of these axial-vector resonances.
}

\section{Conclusions}

We have carried out a theoretical study of the $\tau$ decay into  a pseudoscalar meson plus an axial-vector resonance.
These hadronic decay channels have been very little studied previously, both theoretical and experimentally. Nonetheless, these channels could play an important role in order to shed light on the dynamical formation and structure of the axial-vector resonances. In particlular we  focused on the two lowest mass nonets of axial-vector mesons, $a_1(1260)$
$b_1(1235)$, $h_1(1170)$, $h_1(1380)$, $a_1(1260)$, $f_1(1285)$, and both  poles of the $K_1(1270)$. There has been in the last 15 years compelling  theoretical and experimental evidence that these resonances can be interpreted as molecular or dynamically generated from the interaction of a pseudoscalar and a vector meson in s-wave. Ideed, using the techniques of the chiral unitary approach ($U\chi PT$), which extends the range of aplicability of $\chi PT$ beyond the lowest resonance regions  by the implementation of unitarity in coupled channels to a lowest order amplitude obtained from chiral Lagrangians, poles of the unitarized PV amplitudes were found which could be associated to the known axial-vector resonances. In particular, of great relevance was the prediction that in the strange sector the $K_1(1270)$ actually corresponds to two disctinct poles with different coupling intensities to the different VP channels. Within this framework the dominant production mechanism for the $\tau$ decays considered in the present work is through a triangle mechanism of the kind shown in Fig.~\ref{fig:diane1}, since a vector and a pseudoscalar need to be produced, in addition to the extra final pseudoscalar, to generate the axial-vector resonance.
The initial VV production from the weak current has been theoretically determined, up to a global common factor obtained from the experimental
$\tau \to \nu_\tau K^{*0}K^{*-}$ branching ratio, from a primary $d\bar u$ formation from the $W^-$ boson which then hadronizes producing an extra $q\bar q$ pair within the $^3P_0$ model. The spinor algebra is worked out  following a recent approach where different G-parity contributions could be easily filtered of special interest in the present work.

The pseudoscalar-axial mass distributions predicted in the present work manifest shifts of the strength to the higher energy region of the spectrum partly due to the special shape of the triangle loop function which is carefully evaluated. We make predictions also for integrated branching ratios and, for the only channel experimentally meassured, $\tau^- \to  \nu_\tau \pi^- f_1(1285)$, our result agree with it within uncertainties.
For most of the other channels, the strength of the predicted branching ratios are such that they should be expected to be attainable in experimental studies devoted to exclusive hadronic $\tau$  decays.
 Since the strength of the decays depends crucially on the coupling of the axial-vector resonances to the different VP channels, and these are genuine and non-trivial  predictions of the $U\chi PT$ approach, a positive comparison with those experimental results should reinforce the dynamical or molecular nature of these axial-vector resonances.

\section*{Acknowledgments}

LRD acknowledges the support from the National Natural Science
Foundation of China (Grant No. 11575076) and the State Scholarship Fund of China (No. 201708210057).
This work is partly supported by the Spanish Ministerio
de Economia y Competitividad and European FEDER funds under Contracts No. FIS2017-84038-C2-1-P B
and No. FIS2017-84038-C2-2-P B, and the Generalitat Valenciana in the program Prometeo II-2014/068, and
the project Severo Ochoa of IFIC, SEV-2014-0398 (EO).


\begin{thebibliography}{99}

\bibitem{Schael:2005am}
  S.~Schael {\it et al.} [ALEPH Collaboration],
  Phys.\ Rept.\  {\bf 421} (2005) 191.

\bibitem{Davier:2005xq}
  M.~Davier, A.~Hocker and Z.~Zhang,
  Rev.\ Mod.\ Phys.\  {\bf 78} (2006) 1043.

\bibitem{Braaten:1991qm}
  E.~Braaten, S.~Narison and A.~Pich,
  Nucl.\ Phys.\ B {\bf 373} (1992) 581.


\bibitem{Lafferty:2015hja}
  G.~D.~Lafferty,
  Nucl.\ Part.\ Phys.\ Proc.\  {\bf 260} (2015) 247.




\bibitem{pdg}
  M.~Tanabashi {\it et al.} [Particle Data Group],
  Phys.\ Rev.\ D {\bf 98} (2018) no.3,  030001.



\bibitem{Boito:2014sta}
  D.~Boito, M.~Golterman, K.~Maltman, J.~Osborne and S.~Peris,
  Phys.\ Rev.\ D {\bf 91} (2015) no.3,  034003.



\bibitem{Pich:2013lsa}
  A.~Pich,
  Prog.\ Part.\ Nucl.\ Phys.\  {\bf 75} (2014) 41.


\bibitem{Portoles:2007cx}
  J.~Portoles,
  Nucl.\ Phys.\ Proc.\ Suppl.\  {\bf 169} (2007) 3.


\bibitem{Davier:2013sfa}
  M.~Davier, A.~H枚cker, B.~Malaescu, C.~Z.~Yuan and Z.~Zhang,
  Eur.\ Phys.\ J.\ C {\bf 74} (2014) no.3,  2803.

\bibitem{Lutz:2003fm}
  M.~F.~M.~Lutz and E.~E.~Kolomeitsev,
  Nucl.\ Phys.\ A {\bf 730} (2004) 392.

 \bibitem{roca05}
  L.~Roca, E.~Oset and J.~Singh,   Phys.\ Rev.\ D {\bf 72} 014002, (2005).


\bibitem{Zhou:2014ila}
  Y.~Zhou, X.~L.~Ren, H.~X.~Chen and L.~S.~Geng,
  Phys.\ Rev.\ D {\bf 90} (2014) 014020.


\bibitem{Oset:2018zgc}
  E.~Oset and L.~Roca,
  Phys.\ Lett.\ B {\bf 782} (2018) 332.

\bibitem{landau}
  L.~D.~Landau,   Nucl.\ Phys.\  {\bf 13}, 181 (1959).

 \bibitem{ncol}
 S.~Coleman and R.~E.~Norton,
  Nuovo Cim.\  {\bf 38} (1965) 438.

\bibitem{Bayar:2016ftu}
  M.~Bayar, F.~Aceti, F.~K.~Guo and E.~Oset,
  Phys.\ Rev.\ D {\bf 94} (2016)   074039.

\bibitem{Schmid}
  V.~R.~Debastiani, S.~Sakai and E.~Oset,
  arXiv:1809.06890 [hep-ph].


\bibitem{Dai:2018rra}
  L.~R.~Dai, Q.~X.~Yu and E.~Oset,
  arXiv:1809.11007 [hep-ph].

\bibitem{tdai}
 L.~R.~Dai, R.~Pavao, S.~Sakai and E.~Oset,   arXiv:1805.04573 [hep-ph].


\bibitem{LeYaouanc:1972vsx}
  A.~Le Yaouanc, L.~Oliver, O.~Pene and J.~C.~Raynal,
  Phys.\ Rev.\ D {\bf 8} (1973) 2223.


 \bibitem{close3P0}
F. E. Close. 1979. Academic. An Introduction to Quark and Partons Press.



\bibitem{geng07}
  L. S. Geng, E.~Oset,  L.~Roca,  and  J. A. Oller,   Phys.\ Rev.\ D {\bf 75} 014017, (2007).

\bibitem{micu}
L. Micu, Nucl. Phys. B {\bf 10}, 521 (1969).



\bibitem{bijker}
  E.~Santopinto and R.~Bijker,   Phys.\ Rev.\ C {\bf 82}, 062202 (2010).




 \bibitem{mandl}
 F. Mandl and G. Shaw, Quantum Field Theory, John Wiley \& Sons, 1984.


 \bibitem{sakairamos}
  S.~Sakai, E.~Oset and A.~Ramos,   Eur.\ Phys.\ J.\ A {\bf 54}, 10  (2018).



\bibitem{acetidias} F. Aceti, J. M. Dias and E. Oset,   Eur.\ Phys.\ J.\ A {\bf 51}, 48  (2015).



\end{thebibliography}
\end{document}